\documentclass[aps,eqsecnum,showpacs,amsmath,amssymb,preprint,floatfix]{revtex4}
\usepackage{graphics,epsfig}

\begin{document}


\begin{flushright}
DOE/ER/40762-331\\
UMPP\#05-025
\end{flushright}

\count255=\time\divide\count255 by 60
\xdef\hourmin{\number\count255}
  \multiply\count255 by-60\advance\count255 by\time
 \xdef\hourmin{\hourmin:\ifnum\count255<10 0\fi\the\count255}

\newcommand{\xbf}[1]{\mbox{\boldmath $ #1 $}}

\newcommand{\sixj}[6]{\mbox{$\left\{ \begin{array}{ccc} {#1} & {#2} &
{#3} \\ {#4} & {#5} & {#6} \end{array} \right\}$}}

\newcommand{\threej}[6]{\mbox{$\left( \begin{array}{ccc} {#1} & {#2} &
{#3} \\ {#4} & {#5} & {#6} \end{array} \right)$}}

\title{Pion Photoproduction Amplitude Relations in the $1/N_c$ Expansion}

\author{Thomas D. Cohen}
\email{cohen@physics.umd.edu}

\author{Daniel C. Dakin}
\email{dcdakin@physics.umd.edu}

\affiliation{Department of Physics, University of Maryland,
College Park, MD 20742-4111}

\author{Richard F. Lebed}
\email{Richard.Lebed@asu.edu}

\author{Daniel R. Martin}
\email{daniel.martin@asu.edu}

\affiliation{Department of Physics and Astronomy, Arizona State
University, Tempe, AZ 85287-1504}

\date{December, 2004}

\begin{abstract}
We derive expressions for pion photoproduction amplitudes in the
$1/N_c$ expansion of QCD, and obtain linear relations directly from
this expansion that relate electromagnetic multipole amplitudes at all
energies.  The leading-order relations in $1/N_c$ compare favorably
with available data, while the next-to-leading order relations seem to
provide only a small improvement.  However, when resonance parameters
are compared directly, the agreement at $O(1/N_c)$ or $O(1/N_c^2)$ is
impressive.
\end{abstract}


\maketitle

\section{Introduction \label{sec:intro}}

A recent paper~\cite{CDLN} presented the derivation of linear
relationships among partial-wave amplitudes for $\pi N \! \rightarrow
\! \pi N$ and $\pi N \! \rightarrow \! \pi \Delta$ that hold
in large $N_c$ QCD with only $O(1/N_c^2)$ corrections.  They were
obtained using a model-independent formalism based upon the group
structure of the contracted SU(4) spin-flavor symmetry that emerges in
the single-baryon sector as $N_c \! \rightarrow \! \infty$; this
symmetry by construction ensures consistent $N_c$ power counting for
baryon-meson scattering processes~\cite{GS,DM,DJM}.  The formalism of
Ref.~\cite{CDLN} allows for the inclusion of systematic $1/N_c$
corrections to leading-order results~\cite{CL} among $S$ matrix
elements, specifically the partial-wave amplitudes.  From this
expansion one may obtain linear amplitude relations that hold to
$O(1/N_c^2)$.  As expected, available data support these predictions
better than ones holding only to $O(1/N_c)$~\cite{CDLN}.

The approach for deriving $\pi N$ scattering relations can be applied
to other processes, including single-nucleon Compton scattering,
electron scattering, pion electroproduction ($\gamma^* N \!
\rightarrow \! \pi N$) and photoproduction ($\gamma N \!
\rightarrow \! \pi N$).  In this paper we focus on pion
photoproduction, for which the relevant experimentally accessible
quantities are the electromagnetic multipole amplitudes $M_{L\pm}$ and
$E_{L\pm}$.  We present relations among these multipole amplitudes
that hold to leading order (LO) and next-to-leading order (NLO) in
$1/N_c$~\cite{lead}.

Relations among pion photoproduction amplitudes are not new.  They can
be derived using models in which baryons are considered as chiral
solitons, such as the Skyrme model~\cite{ES,Report}; the
group-theoretical aspects of these models find justification in large
$N_c$ QCD, as discussed in Refs.~\cite{EW,ANW}.  However, the
calculations in Refs.~\cite{ES,Report} do not employ large $N_c$ QCD
as a constraint; the relations obtained there represent a conglomerate
of terms appearing at different orders in $1/N_c$, as discussed in
Sec.~\ref{sec:derive}.  Consequently, the relations in
Refs.~\cite{ES,Report} are not results of the $1/N_c$ expansion.

In this paper, we derive a model-independent expansion for
electromagnetic multipole amplitudes in terms of model-dependent
functions whose coefficients are fixed by group theory.  As shown in
Sec.~\ref{lin}, these model-dependent functions can be algebraically
eliminated to yield seven model-independent linear relations.  These
are compared with experimental data in Sec.~\ref{exp}.  We summarize
in Sec.~\ref{concl}.

We begin by considering general processes of the form $\Phi_1 \! + \!
B_1 \! \rightarrow \! \Phi_2 \! + \! B_2$, where $B_1$ and $B_2$ are
incoming and outgoing nonstrange baryons, and $\Phi_1$ and $\Phi_2$
are incoming and outgoing nonstrange mesons, respectively.  It is also
possible to generalize to scattering processes in which each pair
$B_1$ and $B_2$, and $\Phi_1$ and $\Phi_2$, have a fixed nonzero
strangeness~\cite{CLpenta}.  Amplitude relations for these processes
were first noted in the context of chiral soliton
models~\cite{HEHWMK}, then as model-independent group-theoretical
results derived from a solitonic picture related to the large $N_c$
limit in Refs.~\cite{Mat,MM}, and finally as true model-independent
results of the $1/N_c$ expansion in Ref.~\cite{CL}.  The derivation of
the multipole amplitudes for pion photoproduction is similar to those
of Refs.~\cite{CL,MM}, except that $\Phi_1$ now represents a photon
rather than a meson (or technically, a meson interpolating field with
the quantum numbers of a photon).  Although photons are spin 1, one
precombines the photon spin with its orbital angular momentum relative
to the nucleon target to give the usual multipole angular
momentum~\cite{BLP_RQT} in scattering processes involving radiation.
With this in mind, one begins with the master expression for
meson-baryon partial-wave amplitudes from Refs.~\cite{CL,MM}:
\begin{eqnarray}
S_{L_i L_f S_i S_f I J} & = & \sum_{K, \tilde{K}_i, \tilde{K}_f}
[K] ([R_i][R_f][S_i][S_f][\tilde{K}_i][\tilde{K}_f])^{1/2}
\nonumber \\ & &
\times \left\{ \begin{array}{ccc}
L_i & i_i & \tilde{K}_i \\
S_i & R_i & s_i \\
J & I & K \end{array} \right\} \left\{
\begin{array}{ccc}
L_f & i_f & \tilde{K}_f \\
S_f & R_f & s_f \\
J & I & K
\end{array}
\right\}
\tau_{K \tilde{K}_i \tilde{K}_f L_i L_f},
\label{Mmaster}
\end{eqnarray}
where the \textit{reduced amplitude} $\tau$ is a model-dependent
function that depends only on energy and the quantum numbers $\{K,
\tilde{K}_i, \tilde{K}_f, L_i, L_f\}$.  Its explicit form can be
found only after a particular model of nucleon dynamics, such as
the Skyrme model, is specified.  The notation $[X]$ is shorthand
for $2X \! + \! 1$, the dimension of the spin-$X$ SU(2)
representation.

The quantum numbers specified in Eq.~(\ref{Mmaster}) include the
initial (final) spin=isospin of the nucleon $R_i$ ($R_f$), which
combines vectorially with the initial (final) meson spin $s_i$
($s_f$) to give the total intrinsic spin of the system $S_i$
($S_f$).  These in turn combine with the initial (final)
meson-baryon relative orbital angular momenta $L_i$ ($L_f)$ to
give the total angular momentum $J$. The initial (final) meson
isospin $i_i$ ($i_f$) combines with the nucleon isospin to give
the total isospin $I$.  The effect of constraints from the $1/N_c$
expansion is that the grand spin ${\bf K} \! \equiv \! {\bf I} \!
+ \! {\bf J}$ and the hybrid quantities $\tilde{\bf K}_i \!
\equiv \! {\bf i}_i \! + \! {\bf L}_i$ and $\tilde{\bf K}_f \!
\equiv \! {\bf i}_f \! + \! {\bf L}_f$ provide good quantum
numbers $K$, $\tilde{K}_i$, and $\tilde{K}_f$.  The sums in
Eq.~(\ref{Mmaster}) then run over all values consistent with the
$9j$ symbols, meaning that the entries in each row and column
satisfy a triangle rule.

In using Eq.~(\ref{Mmaster}) to describe the process $\gamma N \!
\rightarrow \! \pi N$, the precombination of photon intrinsic
spin with orbital angular momentum relative to the nucleon target
into a multipole field of order $\ell$ is represented by a simple
mathematical expedient: One sets $s_i \! = \! 0$ and $L_i \! = \!
\ell$ in the first $9j$ symbol.  Since $\ell$ represents the
total of both sources of angular momentum for the photon, the
intrinsic spin of the photon may effectively be set to zero.  As
a side note, the same trick would work for pion
electroproduction, where the photon is virtual and can also
couple through its spin-0 piece.

One important complication must be dealt with before applying
Eq.~(\ref{Mmaster}) to photoproduction processes: The photon has both
isoscalar and isovector pieces.  In large $N_c$ QCD, the leading
isovector coupling of a photon to a ground-state nucleon enters
through the combined spin-flavor operator
\begin{equation} \label{isov}
G^{ia} \equiv \sum_{\alpha=1}^{N_c} \, q^\dagger_\alpha \left(
\frac{\sigma^i}{2} \otimes \frac{\tau^a}{2} \right) q_\alpha \ ,
\end{equation}
where $\sigma$ and $\tau$ are Pauli matrices in spin and isospin
spaces, respectively.  $\alpha$ sums over the $N_c$ quark fields
$q_\alpha$ in the nucleon, but it should be noted that this operator
does not require a quark model to be well defined; in the
field-theoretic context, $q$ simply stands for an interpolating field
with the quantum numbers of a current quark, whose effect summed over
$\alpha$ completely exhausts the full nucleon wave function~\cite{BL}.
In the same language, the (spin-dependent) isoscalar coupling enters
via the operator
\begin{equation}
J^i \equiv \sum_{\alpha=1}^{N_c} \, q^\dagger_\alpha \left(
\frac{\sigma^i}{2} \right) q_\alpha \ .
\end{equation}
The two operators differ in that the matrix elements of the former are
$O(N_c^1)$ for ground-state baryons due to the collective effect of
the $N_c$ quarks, while the matrix elements of the latter are---by
construction---$O(N_c^0)$ for ground-state baryons.  Furthermore,
since the photon couples through the quark charges, it is
straightforward to see that the isovector (isoscalar) couplings have
coefficients $e (q_u \! \mp q_d)$, respectively.  If one takes the
quark charges to have their usual values $q_u \! = \! +\frac 2 3$ and
$q_d \! = \! -\frac 1 3$, as is done in this paper, then the relative
suppression of isoscalar to isovector amplitudes is $1/N_c$.  On the
other hand, if one takes the point of view as in Ref.~\cite{BHL},
where $q_u \! = \! (N_c \! + \! 1)/(2N_c)$ and $q_d \! = \! (1 \! - \!
N_c)/(2N_c)$, then the isoscalar to isovector ratio becomes $1/N_c^2$
(See Ref.~\cite{JJMLM} for a fuller discussion of this point).

Equation~(\ref{Mmaster}) does not manifest this effect.  Because the
anomalous current coupling is suppressed in large $N_c$ due to the
difference in the origin of the isoscalar and isovector pieces, this
feature does not arise in the meson scattering derivation of
Refs.~\cite{CL,MM}.  It must be put in by hand by adding to the
leading isovector ($i_i \! = \! 1$) terms additional isoscalar ($i_i
\! = \! 0$) terms suppressed by an explicit factor $1/N_c$.  This is
purely a feature of isospin breaking in the electromagnetic
interaction: Both isoscalar and isovector couplings couple to the
photon spin.  However, a true spinless isoscalar meson ({\it viz.},
the $\eta$), can couple through the operator
\begin{equation}
\openone \equiv \sum_{\alpha=1}^{N_c} q^\dagger_\alpha q_\alpha \ ,
\end{equation}
whose nucleon matrix elements are $O(N_c^1)$, and therefore couples
just as strongly to nucleons as do pions through the isovector
coupling Eq.~(\ref{isov}).

\section{Derivation \label{sec:derive}}

The derivation of the expression for pion photoproduction multipole
amplitudes begins by substituting $S_i \! = \! S_f \! = \! \frac 1 2$,
$s_i \! = \! s_f \! = \! 0$, $R_i \! = \! R_f \! = \! \frac 1 2$, $i_i
\! \equiv i_\gamma \! \in \! \{0,1\}$ (both of which are of course
added to give the full physical amplitude), and $i_f \! = \! 1$ into
Eq.~(\ref{Mmaster}):
\begin{eqnarray}
S_{\ell L \frac{1}{2} \frac{1}{2} I J} =
2 (-1)^{L-\ell} \sum_K [K] \left\{ \begin{array}{ccc}
J & \ell & \frac{1}{2} \\
i_\gamma & I & K \\
\end{array} \right\}
\left\{ \begin{array}{ccc}
J & L & \frac{1}{2} \\
i_f & I & K \\
\end{array} \right\}
\tau_{K \ell L} \ . \label{RedMast}
\end{eqnarray}

From Eq.~(\ref{RedMast}) one obtains the form of the multipole
amplitudes by including the isospin Clebsch-Gordan coefficients
specifying the initial and final nucleon charge states.  Using $\nu$
for the pion isospin third component and $m_I$ for that of the
incoming nucleon, the multipole amplitude for a specific charge
channel is
\begin{eqnarray}
M_{\ell L J m_I \nu}^{\lambda I i_\gamma} & \! \! \! = \! \!
& \left( \! \! \!
{\begin{array}{cc} {\begin{array}{cc} 1 & \frac{1}{2} \\ \nu & m_I \!
- \! \nu \end{array}} & \! \! \!  {\left| {\begin{array}{c} I \\ m_I
\\ \end{array}} \right.} \\ \end{array}} \! \! \!
\right) \! \! \left( \! \! \! {\begin{array}{cc}
 {\begin{array}{cc}
   i_\gamma & \frac{1}{2}  \\
   0 & m_I  \\
  \end{array}} & \! \! \!
{\left|
 {\begin{array}{c}
   I  \\
   m_I  \\
  \end{array}} \right.} \\
 \end{array}} \! \! \!
\right)
\! 2 (-1)^{L-\ell} \sum_K [K] \! \left\{
\begin{array}{ccc}
J & \ell & \frac{1}{2} \\
i_\gamma & I & K
\end{array}
\right\} \! \left\{
\begin{array}{ccc}
J & L & \frac{1}{2} \\
1 & I & K
\end{array}
\right\} \tau_{K \ell L}^{\lambda} \, .
\nonumber \\ & & \label{RedMmast}
\end{eqnarray}
The index $\lambda$ indicates the type of multipole, and is determined
by the relative parity of $\ell$ and $L$: ($\ell \! - \! L$) odd gives
electric (e) multipoles, ($\ell \! - \! L$) even gives magnetic (m)
multipoles.

This expansion is most useful when written in terms of $t$-channel
exchange amplitudes, since large $N_c$ QCD restricts their form as
discussed in Refs.~\cite{CDLN,KapMan}: The leading amplitudes in
$1/N_c$ have $I_t \! = \! J_t$~\cite{MM}, and the amplitudes with
$|I_t \! - \! J_t| \! = \! n$ are suppressed by a relative factor
$1/N_c^n$.  Following Ref.~\cite{CDLN}, we compute the $t$-channel
amplitudes for the separate cases where $i_\gamma \! = \! 1$ and
$i_\gamma \! = \! 0$.  Using the Biedenharn-Elliot sum
rule~\cite{edmonds}, one can rewrite the product of $6j$ symbols in
Eq.~(\ref{RedMmast}) as
\begin{eqnarray}
\left\{
\begin{array}{ccc}
  J & \ell & \frac{1}{2} \\
  i_\gamma & I & K
\end{array}
\right\} \left\{
\begin{array}{ccc}
  J & L & \frac{1}{2} \\
  i_f & I & K
\end{array}
\right\} & = & \sum_{{\cal J}} \frac{(-1)^{2{\cal
J}-i_f+i_\gamma}[{\cal J}]} {2\sqrt{[i_f][i_\gamma][L][\ell]}}
\left[
\begin{array}{ccc}
  i_f & L & K \\ \ell & i_\gamma & {\cal J}
\end{array}
\right] \left[
\begin{array}{ccc}
  i_f & \frac{1}{2} & I \\
  \frac{1}{2} & i_\gamma & {\cal J}
\end{array}
\right] \left[
\begin{array}{ccc}
  L & \frac{1}{2} & J \\
  \frac{1}{2} & \ell & {\cal J}
\end{array}
\right] \ , \label{BeidEll}
\nonumber \\ & &
\end{eqnarray}
where the modified $6j$ symbols (called $[6j]$ symbols in
Ref.~\cite{CDLN}) are defined by
\begin{eqnarray}
\left\{
\begin{array}{ccc}
  a & b & e \\
  c & d & f
\end{array}
\right\} \equiv
\frac{(-1)^{-(b+d+e+f)}}{([a][b][c][d])^{\frac{1}{4}}} \left[
\begin{array}{ccc}
  a & b & e \\
  c & d & f
\end{array}
\right] \ . \label{Red6j}
\end{eqnarray}
Note that the $[6j]$ and the usual $6j$ symbols share the same
triangle rules.

The full $t$-channel multipole amplitude can now be written in terms
of $[6j]$ symbols, using Eqs.~(\ref{RedMmast}) and (\ref{BeidEll})
with $i_f \! = \! 1$ ($[1] \! \rightarrow \!  3$) for the pion.  It is
convenient to define $t$-channel amplitudes by
\begin{eqnarray}
\tau_{{\cal J} \ell L}^{t \lambda i_\gamma} & \equiv &
\frac{(-1)^{2 {\cal J} -1 + i_\gamma} [{\cal J}]}
{\sqrt{[1][i_\gamma][L][\ell]}}
\sum_{K} \, [K] \left[
\begin{array}{ccc}
     1 & L & K \\ \ell & i_\gamma & {\cal J}
\end{array}
\right] \tau_{K \ell L}^{\lambda} \ . \label{t_chan}
\end{eqnarray}
Then, for the isovector case ($i_\gamma \!  = \! 1$),
\begin{eqnarray}
\hspace{-1em} M_{\ell L J m_I \nu}^{\lambda I 1} & \! = \! &
(-1)^{L-\ell}
\left( \!
\! \! {\begin{array}{cc}
 {\begin{array}{cc}
    1 & \frac{1}{2}  \\
    \nu & m_I \! - \! \nu
  \end{array}} & \! \! \!
{\left|
 {\begin{array}{c}
    I \\
    m_I \\
  \end{array}} \right.} \\
 \end{array}} \! \! \!
\right) \! \! \left( \! \! \! {\begin{array}{cc}
 {\begin{array}{cc}
   1 & \frac{1}{2}  \\
   0 & m_I  \\
  \end{array}} & \! \! \!
{\left|
 {\begin{array}{c}
   I  \\
   m_I  \\
  \end{array}} \right.} \\
 \end{array}} \! \! \!
\right) \! \sum_{{\cal J}} \left[
\begin{array}{ccc}
  1 & \frac{1}{2} & I \\
  \frac{1}{2} & 1 & {\cal J}
\end{array}
\right] \! \! \left[
\begin{array}{ccc}
  L & \frac{1}{2} & J \\
  \frac{1}{2} & \ell & {\cal J}
\end{array}
\right] \tau_{{\cal J} \ell L}^{t \lambda 1} \label{RedMmast_i1} \, .
\end{eqnarray}
In the isoscalar case ($i_\gamma \! = \! 0$), the first $6j$ symbol
and the second Clebsch-Gordan coefficient in Eq.~(\ref{RedMmast})
collapse to simple factors times Kronecker $\delta$'s.  Including the
explicit $1/N_c$ suppression described above, one has
\begin{eqnarray}
M_{\ell L J m_I \nu}^{\lambda I 0} & = &
\frac{(-1)^{L-\ell}}{N_c} \left( \! \! \! {\begin{array}{cc}
 {\begin{array}{cc}
    1 & \frac{1}{2}  \\
    \nu & m_I \! - \! \nu
  \end{array}} & \! \! \!
{\left|
 {\begin{array}{c}
    \frac{1}{2} \\
    m_I \\
  \end{array}} \right.} \\
 \end{array}} \! \! \!
\right) \frac{\delta_{I,\frac{1}{2}}}{[1]^{1/4}} \left[
\begin{array}{ccc}
  L & \frac{1}{2} & J \\
  \frac{1}{2} & \ell & 1
\end{array}
\right] \tau_{1 \ell L}^{t \lambda 0} \ . \label{RedMmast_i0}
\end{eqnarray}

Equations (\ref{RedMmast_i1}) and (\ref{RedMmast_i0}) are
analogous to expressions obtained earlier by Eckart and
Schwesinger~\cite{ES}, if one identifies $\tau_{K \ell
L}(i_\gamma \! = \! 1)$ and $\tau_{K \ell L}(i_\gamma \! = \! 0)$
with their dynamical functions $V_{K \ell}^L(k_\gamma,k_\pi)$ and
$S_{K \ell}^L(k_\gamma,k_\pi)$, respectively. Reference~\cite{ES}
studied photoproduction of baryon resonances in the context of
the Skyrme model and derived expressions for the same multipole
amplitudes as considered here.  Their expansion, however,
includes a third dynamical function $R_{K
\ell}^L(k_\gamma,k_\pi)$ that does not appear in the present
derivation, since it represents Skyrmion angular velocity terms,
which vanish at leading order in $1/N_c$.  Similarly, our
analysis would suppress $S$ compared to $V$ by the aforementioned
$1/N_c$ factor. The linear relations derived in Refs.~\cite{ES}
and \cite{Report} are therefore not consequences of large $N_c$
QCD since all the functions $R$, $S$, and $V$ are treated as
equally important in their analysis.

Because of the extra $1/N_c$ suppression of isoscalar compared to
isovector amplitudes, an expansion to consistent order in $1/N_c$
requires the inclusion of NLO amplitudes for just the isovector
channel.  We parameterize them by following the same procedure as
in Ref.~\cite{CDLN}: The LO terms all have $|I_t \! - \! J_t| \!
= \! 0$~\cite{MM}, while all linearly independent NLO terms have
$|I_t \! = \! J_t| \! = \! 1$~\cite{CDLN,KapMan}.  Generalizing
Eq.~(\ref{RedMmast_i1}) in this way gives
\begin{eqnarray}
M_{\ell L J m_I \nu}^{\lambda I 1 \, (\textrm{NLO})} & \! = &
\frac{(-1)^{L-\ell}}{N_c} \left( \! \! \! {\begin{array}{cc}
 {\begin{array}{cc}
    1 & \frac{1}{2}  \\
    \nu & m_I \! - \! \nu
  \end{array}} & \! \! \!
{\left|
 {\begin{array}{c}
    I \\
    m_I \\
  \end{array}} \right.} \\
 \end{array}} \! \! \!
\right) \left( \! \! \! {\begin{array}{cc}
 {\begin{array}{cc}
    1 & \frac{1}{2}  \\
    0 & m_I
  \end{array}} & \! \! \!
{\left|
 {\begin{array}{c}
    I \\
    m_I \\
  \end{array}} \right.} \\
 \end{array}} \! \! \!
\right) \nonumber
\\ & & \times
\left\{
\sum_{x} \left[
\begin{array}{ccc}
  1 & \frac{1}{2} & I \\
  \frac{1}{2} & 1 & x
\end{array}
\right] \! \! \left[
\begin{array}{ccc}
  L & \frac{1}{2} & J \\
  \frac{1}{2} & \ell & x \! + \! 1
\end{array}
\right] \! \tau_{x \ell L}^{t \lambda (+)} + \sum_{y} \left[
\begin{array}{ccc}
  1 & \frac{1}{2} & I \\
  \frac{1}{2} & 1 & y
\end{array}
\right] \! \! \left[
\begin{array}{ccc}
  L & \frac{1}{2} & J \\
  \frac{1}{2} & \ell & y \! - \! 1
\end{array}
\right] \! \tau_{y \ell L}^{t \lambda (-)}
\right\}. \nonumber \\ & & \label{RedMmast_NLO}
\end{eqnarray}
Note that ${\cal J}$ in Eq.~(\ref{RedMmast_i1}), and $x$ and $y$ in
Eq.~(\ref{RedMmast_NLO}), are dummy labels for $I_t$ in each
corresponding sum.  The total multipole amplitude expansion, including
all LO terms and good to consistent order in $1/N_c$, is the sum of
Eqs.~(\ref{RedMmast_i1}), (\ref{RedMmast_i0}), and
(\ref{RedMmast_NLO}):
\begin{eqnarray}
M_{\ell L J m_I \nu}^{\lambda I} & = & M_{\ell L J m_I
\nu}^{\lambda I 1} + M_{\ell L J m_I \nu}^{\lambda I 0} + M_{\ell
L J m_I \nu}^{\lambda I 1 \, (\textrm{NLO})}. \label{TotAmp}
\end{eqnarray}

Some general comments apply to Eq.~(\ref{TotAmp}).  First, the NLO
amplitude contains only two sums since $|I_t \! - \! J_t|=1$ is
satisfied only by $I_t \! = \! J_t \! \pm \! 1$.  The sum over ${\cal
J}$ is constrained to $0$ and $1$, due to the triangle rule
$\Delta(\frac{1}{2},\frac{1}{2},{\cal J})$ in the first $[6j]$ symbol
in Eq.~(\ref{RedMmast_i1}).  Similarly, triangle rules in the second
$[6j]$ symbol reduce the respective sums over $x$ and $y$ to single
terms with $x \! = \! 0$ and $y \! = \! 1$.  Equation~(\ref{TotAmp})
in its simplest expanded form, with free quantum numbers $J$, $L$,
$\ell$, $m_I$, and $\nu$, reads
\begin{eqnarray}
M_{\ell L J}^{\lambda m_I \nu} & = & \sum_{I}
(-1)^{L-\ell} \left(
\begin{array}{cc}
   1 & \frac{1}{2}  \\
   \nu & m_I \! - \! \nu  \\
  \end{array}
\right| \left.
\begin{array}{c}
   I  \\
   m_I  \\
  \end{array}
\right) \nonumber \\ & \times &
\left[ \left(
   \begin{array}{cc}
   1 & \frac{1}{2}  \\
   0 & m_I  \\
  \end{array}
\right| \left.
   \begin{array}{c}
   I  \\
   m_I  \\
  \end{array} \right) \right.
\left\{ \delta_{\ell , L} \tau_{0 L L}^{t \lambda 1} +
\sqrt{\frac{2}{3}} \left( \delta_{I,\frac{1}{2}}-\frac{1}{2}
\delta_{I,\frac{3}{2}} \right)
\left[
\begin{array}{ccc}
  L & \frac{1}{2} & J \\
  \frac{1}{2} & \ell & 1
\end{array}
\right] \tau_{1 \ell L}^{t \lambda 1}
\right. \nonumber \\ & &
\left. \left.
+ \frac{1}{N_c} \left( \left[
\begin{array}{ccc}
  L & \frac{1}{2} & J \\
  \frac{1}{2} & \ell & 1
\end{array}
\right] \tau_{0 \ell L}^{t \lambda (+)} + \sqrt{\frac{2}{3}}
\left( \delta_{I,\frac{1}{2}}-\frac{1}{2} \delta_{I,\frac{3}{2}}
\right) \delta_{\ell,L} \tau_{1 L L}^{t \lambda (-)} \right)
\right\} \right. \nonumber \\ & & \left.
+ \frac{1}{N_c} \frac{\delta_{I,\frac{1}{2}}}{[1]^{1/4}}
\left[
\begin{array}{ccc}
  L & \frac{1}{2} & J \\
  \frac{1}{2} & \ell & 1
\end{array}
\right] \tau_{1 \ell L}^{t \lambda 0} \right] .
\label{Final}
\end{eqnarray}

\section{Linear Relations \label{lin}}

Charge conservation limits the number of pion photoproduction channels
to four: $\gamma p \rightarrow \pi^+n$, $\gamma n
\rightarrow \pi^-p$, $\gamma p \rightarrow \pi^0p$, $\gamma n
\rightarrow \pi^0n$.  However, due to isospin invariance of the
strong interaction, only three of these are independent.  Since the
species in $\gamma n \rightarrow \pi^0n$ are neutral, this reaction is
difficult to study experimentally and we use isospin freedom to
eliminate its amplitudes (separately for total $I \! = \! \frac 1 2$
and $\frac 3 2$ channels)~\cite{neut}.  The remaining three charged
channels can occur via an electric or magnetic transition.  In the
magnetic case, the photon and pion have the same orbital angular
momentum ($\ell \! = \! L$), whereas in the electric case, there is a
change of one unit ($\ell \! = \! L \! \pm \! 1$).  Given these
restrictions, the set of multipole amplitudes describing these cases
can be written in terms of a still {\it smaller} set of reduced
amplitudes.  Thus one expects linear relations among the physically
measurable amplitudes.

Linear relations can be derived at both LO and NLO in $1/N_c$.  In
order to find the LO relations, we work with only the LO pieces in
Eq.~(\ref{Final}) (\textit{i.e.}, disregard the $1/N_c$-suppressed
terms).  To find the relations that hold to NLO, we use the complete
expression.  The electric and magnetic transitions have distinct
expansions and are investigated independently.

Let us begin with the expansion of the electric multipole amplitudes.
Six physical amplitudes correspond to the two ways, $J \! = \! L \! \pm
\! \frac 1 2$, of combining the pion and nucleon angular momenta for
each of the three charged reactions.  At LO these are expanded in
terms of only two reduced amplitudes ($\tau^{t \, \textrm{e} \, 1}_{1,
L \pm 1, L}$), implying four relations.  Two of these are:
\begin{eqnarray}
M^{\textrm{e},\, p(\pi^+)n}_{L-1,L,-}=M^{\textrm{e},\,
n(\pi^-)p}_{L-1,L,-}+O(N_c^{-1}) \ (L \geq 2),\label{LOEminus} \\
M^{\textrm{e},\, p(\pi^+)n}_{L+1,L,+}=M^{\textrm{e},\,
n(\pi^-)p}_{L+1,L,+}+O(N_c^{-1}) \ (L \geq 0) , \label{LOEplus}
\end{eqnarray}
where the last subscript in each amplitude is no longer $J$, but
represents the equivalent information of the sign in $J
\! = \! L \! \pm \! \frac 1 2$.  These relations follow
simply from isospin symmetry of isovector amplitudes, since the
isoscalar component of the photon current is absent at LO.

The other two LO relations imply the vanishing of the electric
multipole amplitudes for $\gamma p \rightarrow \pi^0 p$ at leading
order in $1/N_c \,$:
\begin{equation} \label{pi0e}
M_{L\pm1,L,\pm}^{\textrm{e},\, p(\pi^0)p} = O(1/N_c) \ .
\end{equation}
After extrapolating to the real world of $N_c \! = \! 3$, one expects
these amplitudes to be about a factor $N_c \! = \! 3$ smaller (on
average) than those of the charge-exchange reactions.

Once the NLO terms in Eq.~(\ref{Final}) are included, four new reduced
amplitudes ($\tau^{t \, \textrm{e} (+)}_{0, L \pm 1, L}$ and $\tau^{t
\, \textrm{e} \, 0}_{1, L \pm 1 , L}$) appear, leaving no remaining
electric multipole relations at NLO.

Turning to the magnetic transition, one sees that only two LO reduced
amplitudes ($\tau^{t \, \textrm{m} \, 1}_{1 L L}$ and $\tau^{t \,
\textrm{m} \, 1}_{0 L L}$) are needed to describe the six physical
amplitudes.  This yields four LO linear relations:
\begin{eqnarray}
M^{\textrm{m},\,p(\pi^0)p}_{L,L,-}
& = &
M^{\textrm{m},\,p(\pi^0)p}_{L,L,+}+O(N_c^{-1}) \ (L \geq 1),
\label{LOMnought} \\
M^{\textrm{m},\,p(\pi^+)n}_{L,L,-}
& = &
M^{\textrm{m},\,n(\pi^-)p}_{L,L,-} =
-\frac{L+1}{L}M^{\textrm{m},\,p(\pi^+)n}_{L,L,+} =
-\frac{L+1}{L}M^{\textrm{m},\,n(\pi^-)p}_{L,L,+} \ (L \geq 1).
\label{LOMchain}
\end{eqnarray}
As before, two of these follow from isospin symmetry among the
isovector amplitudes.  The NLO terms bring in only three more reduced
amplitudes ($\tau^{t \, \textrm{m} (+)}_{0 L L}$, $\tau^{t \,
\textrm{m} (-)}_{1 L L}$, and $\tau^{t \, \textrm{m} \, 0}_{1 L L}$),
meaning that one relation remains at this order.  Indeed, one might
have anticipated fewer amplitudes in the magnetic rather than the
electric transition since, in the former case, only $\ell \! = \! L$
is allowed.  The NLO relation is
\begin{equation}
M^{\textrm{m},\,p(\pi^+)n}_{L,L,-}
=
M^{\textrm{m},\,n(\pi^-)p}_{L,L,-}
-\left(\frac{L+1}{L}\right)
\left[
M^{\textrm{m},\,p(\pi^+)n}_{L,L,+}-M^{\textrm{m},\,n(\pi^-)p}_{L,L,+}
\right]
+O(N_c^{-2}) \ (L\geq 1).\label{NLO}
\end{equation}
A casual glance shows this to be a linear combination of the LO
relations in Eq.~(\ref{LOMchain}); it is the unique combination for
which the NLO corrections (in brackets) cancel as well.  One expects
this relation to hold empirically a factor of $N_c \! = \! 3$ better
than its LO counterpart.

Finally, we point out that a number of relations may be obtained from
Eq.~(\ref{Final}) for pure $I \! = \! \frac 1 2$ or $\frac 3 2$
amplitude combinations, but this merely represents a different basis
for representing the charge states.

\section{Experimental Tests \label{exp}}

In principle, all seven linear relations included in
Eqs.~(\ref{LOEminus})--(\ref{NLO}) (for each allowed value of $L$),
plus the smallness of Eqs.~(\ref{pi0e}), can be tested by comparison
with available experimental data.  The numbers used are the results of
partial-wave analysis applied to raw data from experiments in which
real photons are scattered off nucleon targets.  We use the data
presented by the SAID program~\cite{SAID} at George Washington
University and the MAID~2003 program~\cite{MAID} at Universit\"{a}t
Mainz.  Although one requires only a single data set, it is useful to
check the extent to which the model dependence of the data analysis
used by the two groups affects our comparisons.  We find that the
difference is not significant for our tests.

It is now convenient to introduce the notation used in the
experimental data tables.  The \textit{electromagnetic multipoles} are
given in terms of our multipole amplitudes:
\begin{eqnarray}
M^{\textrm{e}}_{L-1,L,-} & = & +\beta \sqrt{L(L-1)}\, E_{L-}
\nonumber \\
M^{\textrm{e}}_{L+1,L,+} & = & +\beta \sqrt{(L+2)(L+1)} \, E_{L+}
\nonumber \\
M^{\textrm{m}}_{L,L,+} & = & -\beta \sqrt{L(L+1)} \, M_{L+}
\nonumber \\
M^{\textrm{m}}_{L,L,-} & = & -\beta \sqrt{L(L+1)}\, M_{L-}
\label{multipoles}
\end{eqnarray}
where~\cite{ES}
\begin{equation}
\beta \equiv -F_\pi \sqrt{ \frac{k_\gamma}{8\pi\alpha} } \ ,
\end{equation}
with $F_\pi \simeq 186 \, {\rm MeV}$ and $k_\gamma$ the photon c.m.\
3-momentum, is an energy scale that cancels from all linear relations
and therefore is irrelevant to this work: From Sec.~\ref{lin}, one
notes that each term in any one of our linear relations has the same
prefactors of $\beta$ and $L$ entering via Eq.~(\ref{multipoles}).
The relations therefore take the same form when written in terms of
the electromagnetic multipoles.  It should also be noted that the
convention for the signs of $p (\pi^+) n$ amplitudes appearing in data
are often reversed (in MAID, for example) compared to those fixed by
the standard Condon-Shortley convention used in this paper.


In all plots we present both real and imaginary parts of partial wave
amplitudes, for values of c.m.\ energy $W$ of the $\gamma N$ system up
to 2~GeV.

\begin{figure}[ht]
\epsfxsize 2.5 in \epsfbox{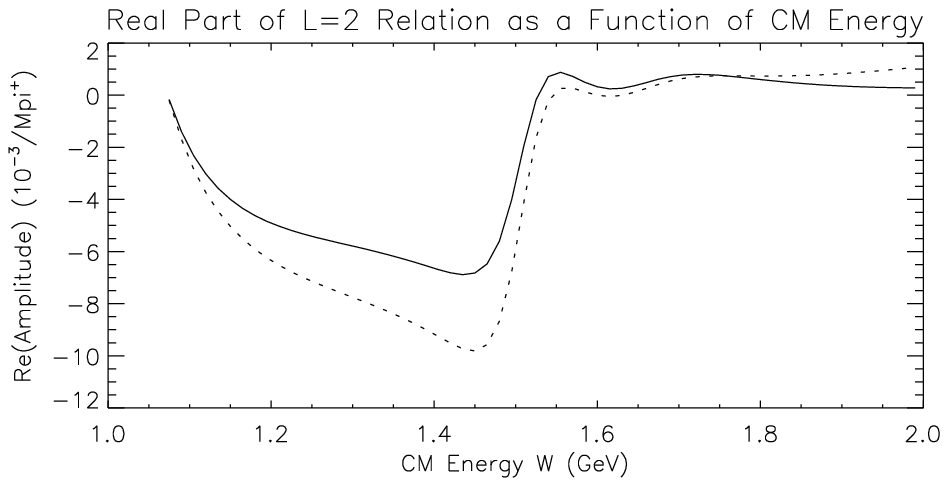} \hspace{5em}
\epsfxsize 2.5 in \epsfbox{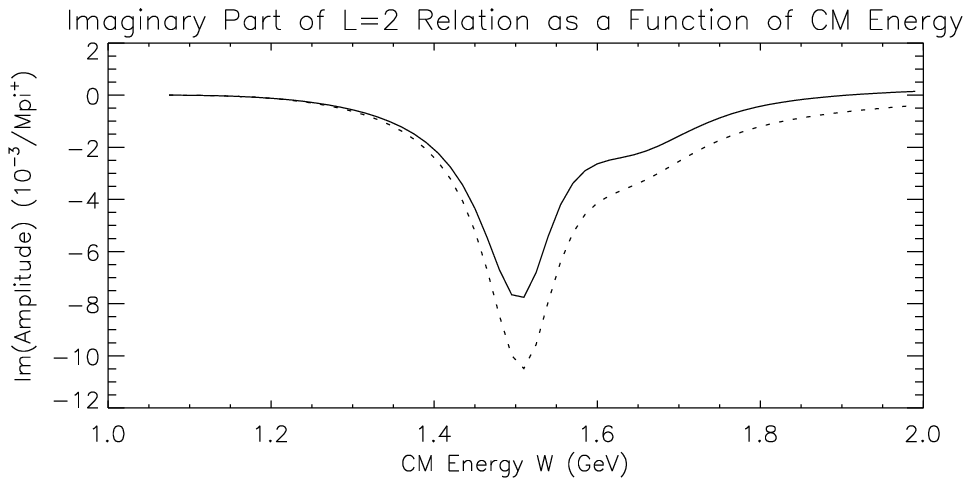} \\
\epsfxsize 2.5 in \epsfbox{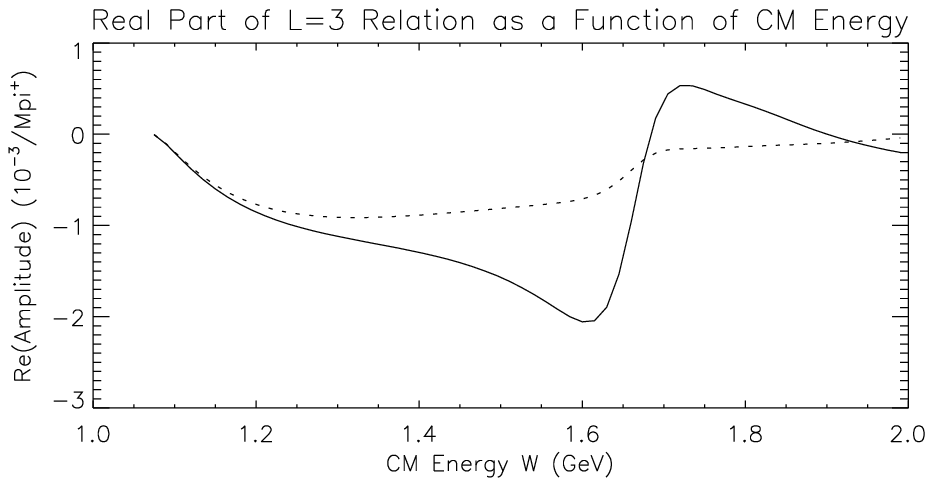} \hspace{5em}
\epsfxsize 2.5 in \epsfbox{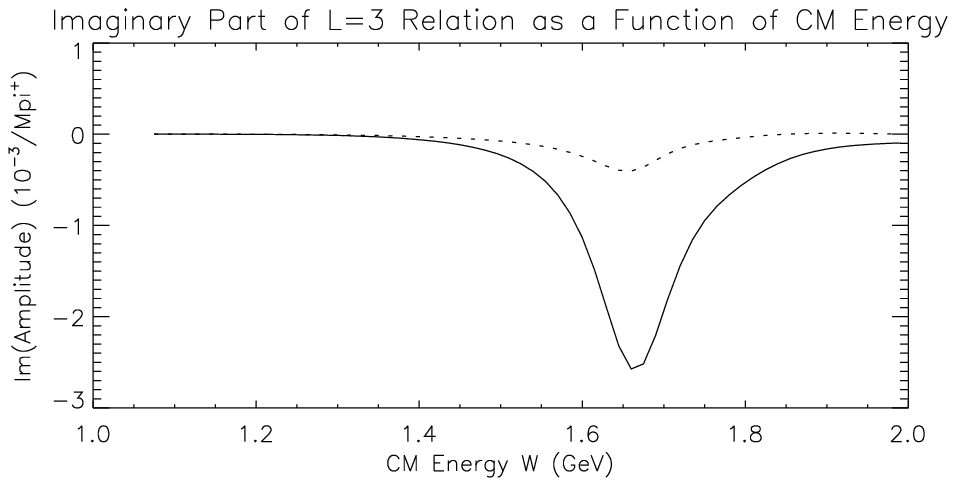} \\
\epsfxsize 2.5 in \epsfbox{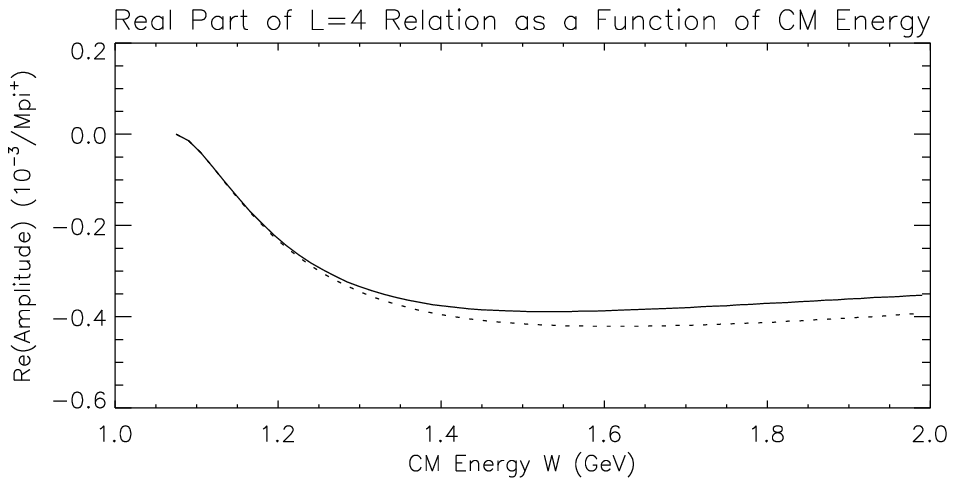} \hspace{5em}
\epsfxsize 2.5 in \epsfbox{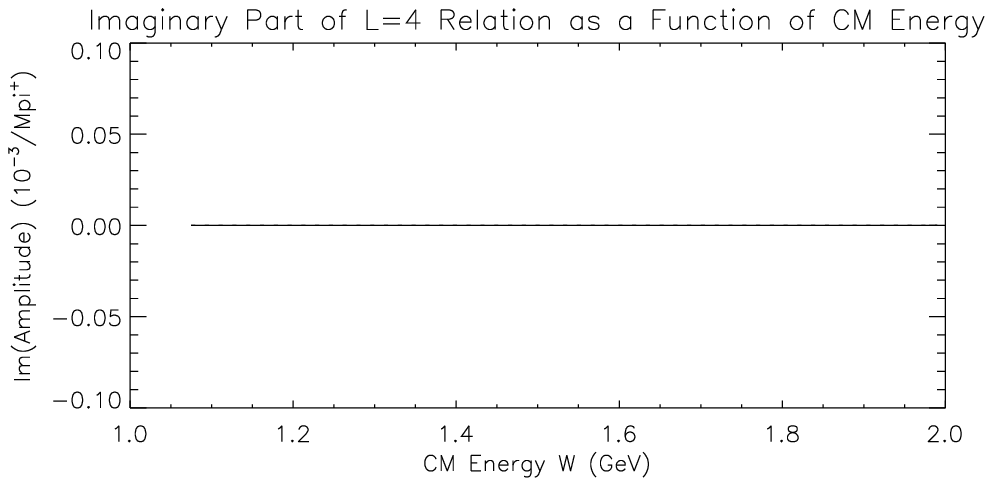} \\
\epsfxsize 2.5 in \epsfbox{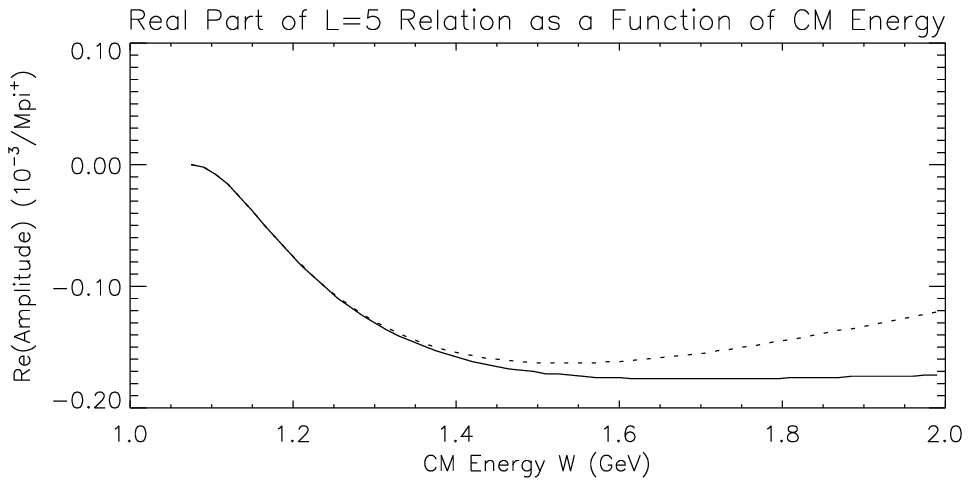} \hspace{5em}
\epsfxsize 2.5 in \epsfbox{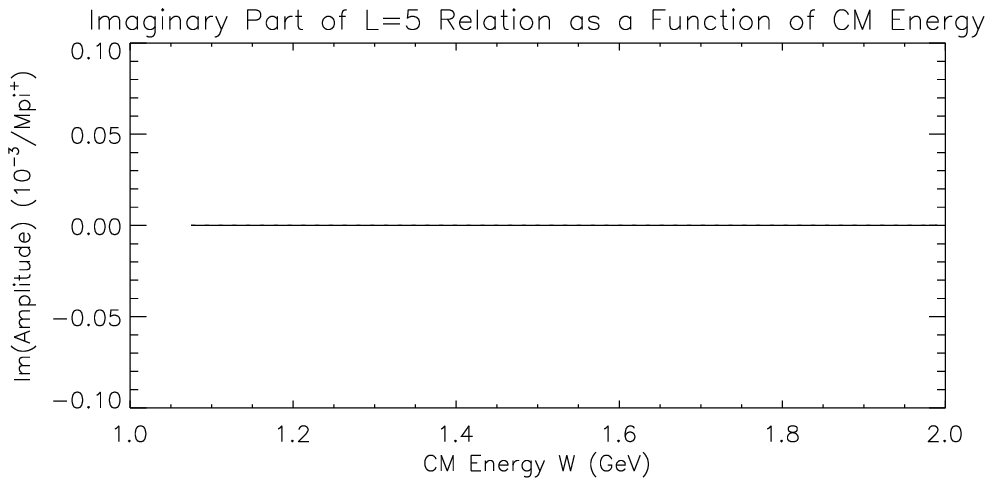}
\caption{
Electric multipole data from the MAID~2003 website~\cite{MAID}.  Solid
lines indicate the l.h.s.\ of relation~(\ref{LOEminus}) for values $L
\! \ge \! 2$, while dotted lines represent the r.h.s.}
\label{LOEmplot}
\end{figure}

\begin{figure}[ht]
\epsfxsize 2.5 in \epsfbox{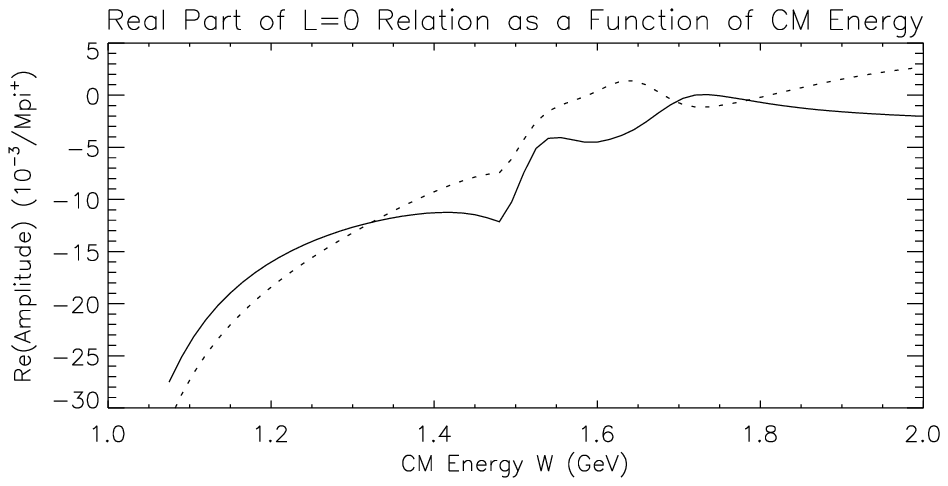} \hspace{5em}
\epsfxsize 2.5 in \epsfbox{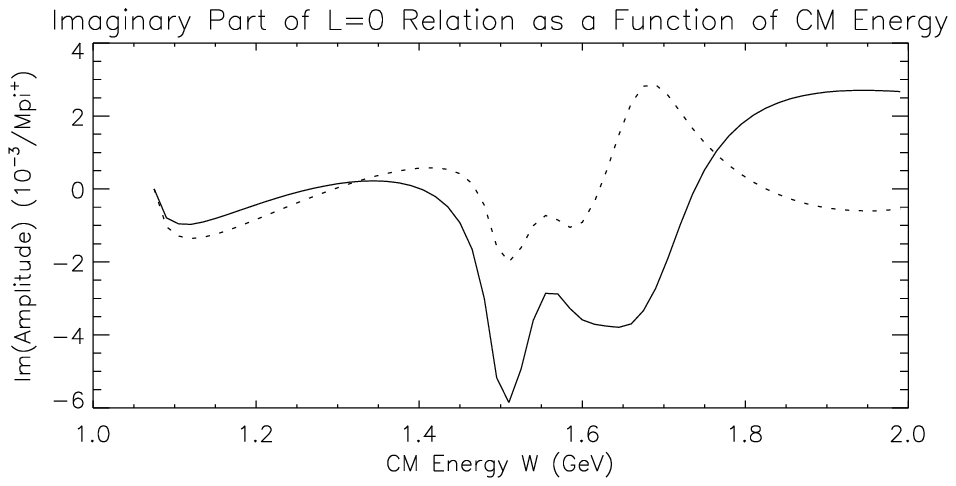} \\
\epsfxsize 2.5 in \epsfbox{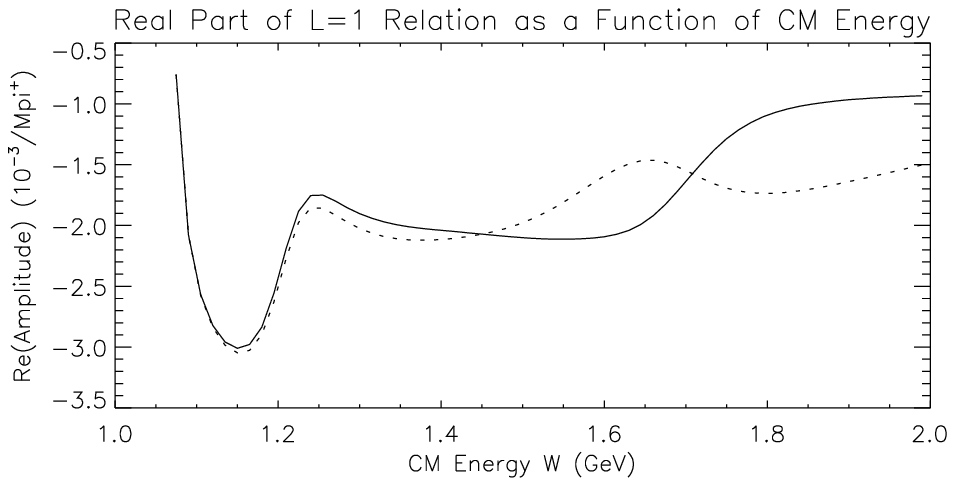} \hspace{5em}
\epsfxsize 2.5 in \epsfbox{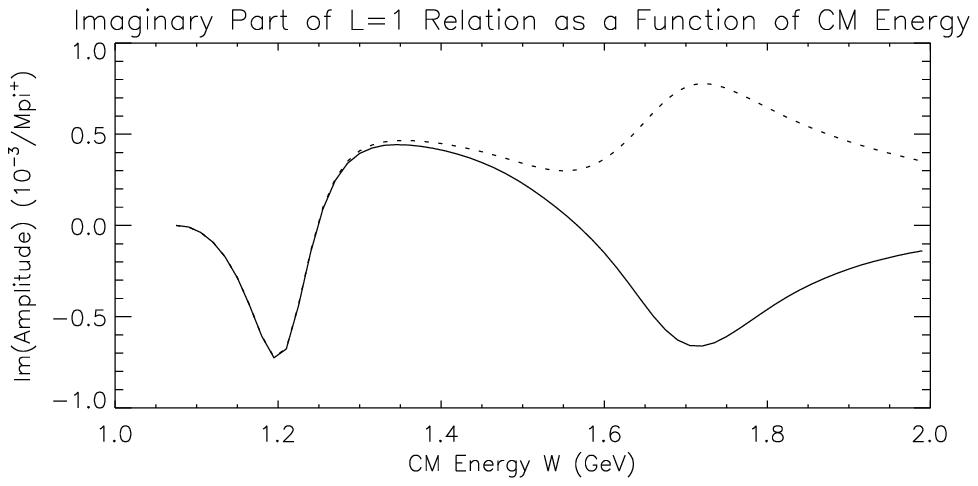} \\
\epsfxsize 2.5 in \epsfbox{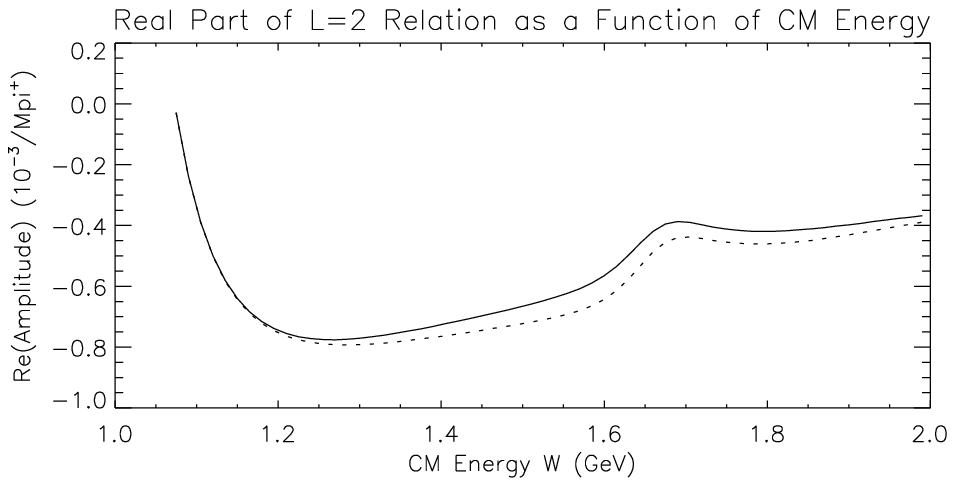} \hspace{5em}
\epsfxsize 2.5 in \epsfbox{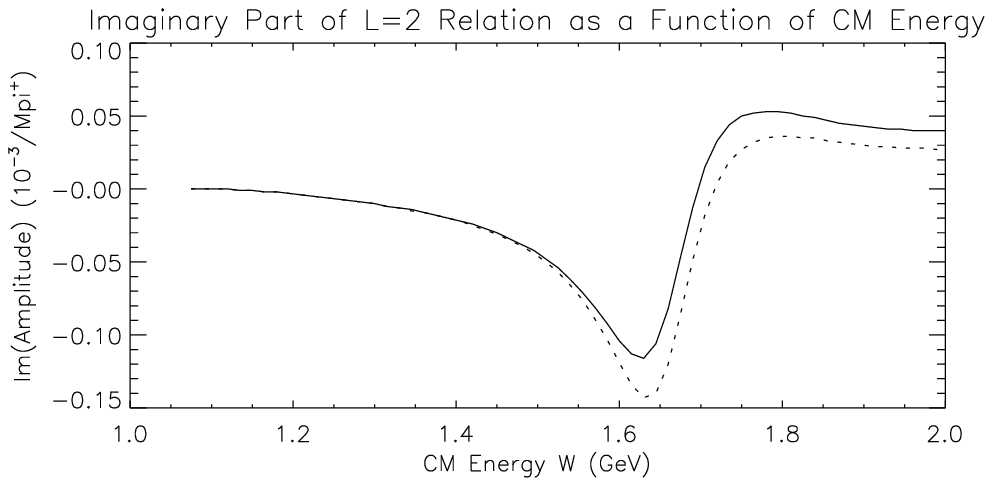} \\
\epsfxsize 2.5 in \epsfbox{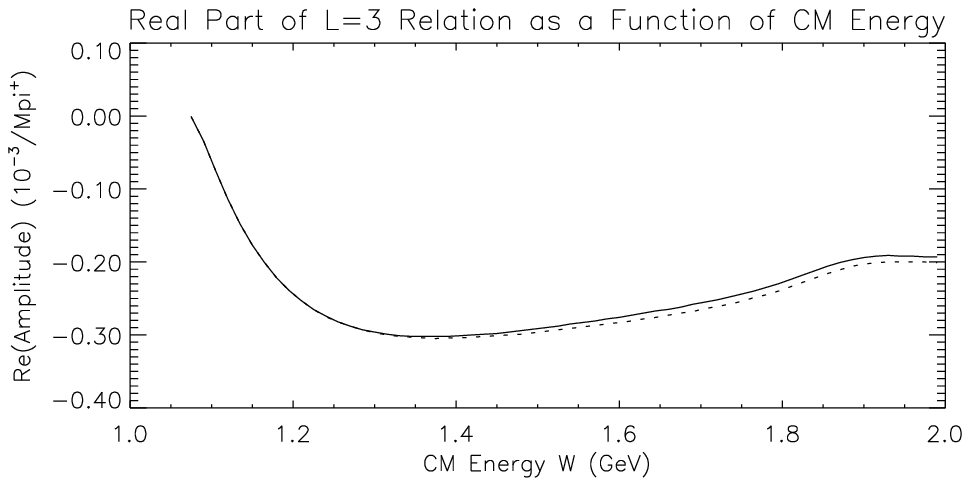} \hspace{5em}
\epsfxsize 2.5 in \epsfbox{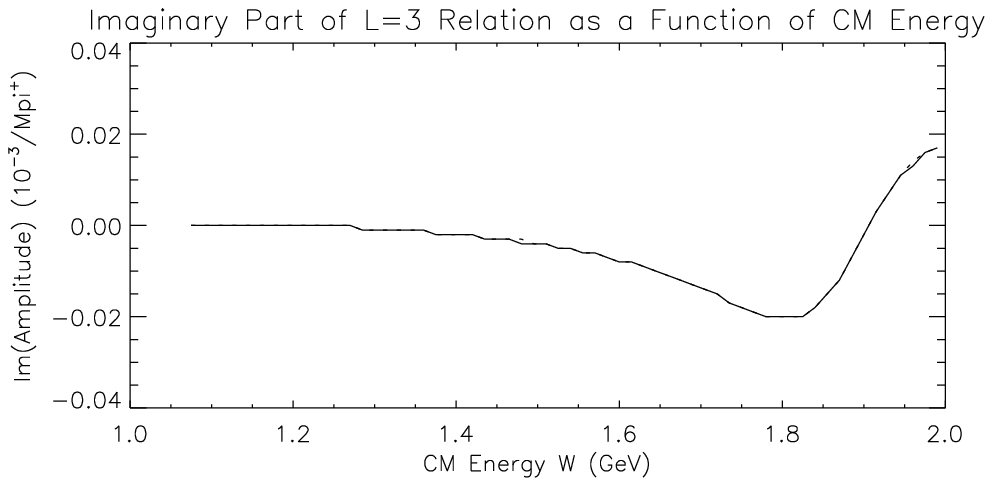} \\
\epsfxsize 2.5 in \epsfbox{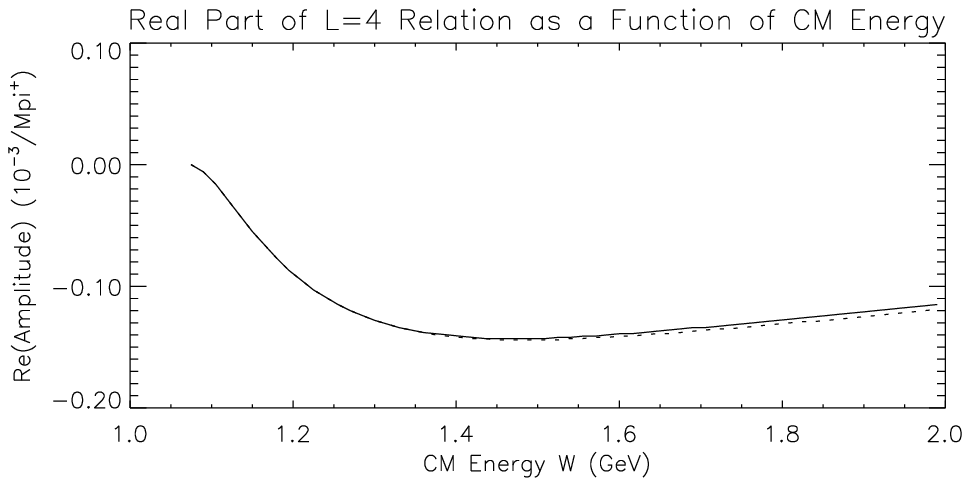} \hspace{5em}
\epsfxsize 2.5 in \epsfbox{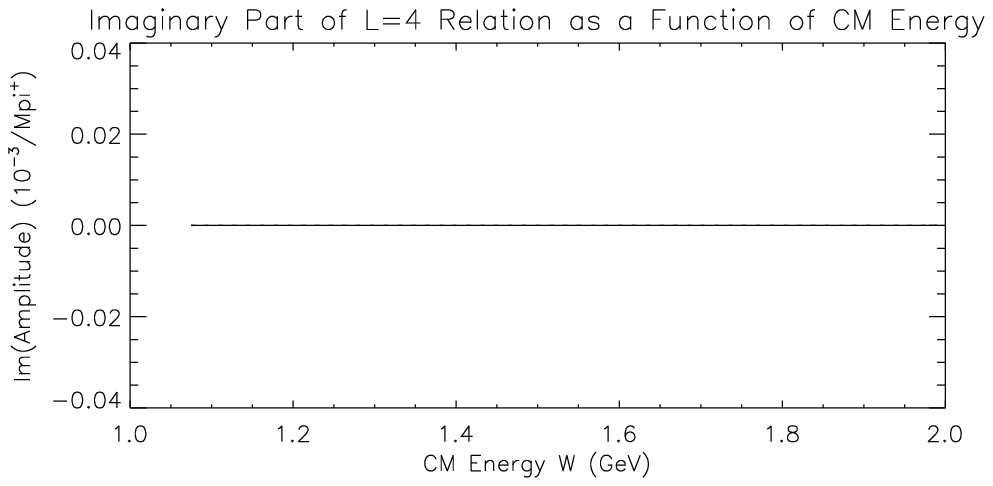} \\
\epsfxsize 2.5 in \epsfbox{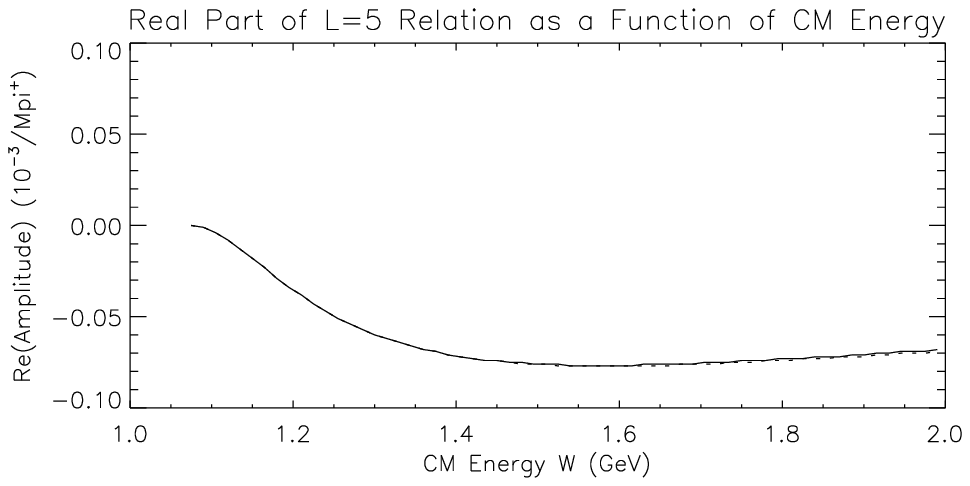} \hspace{5em}
\epsfxsize 2.5 in \epsfbox{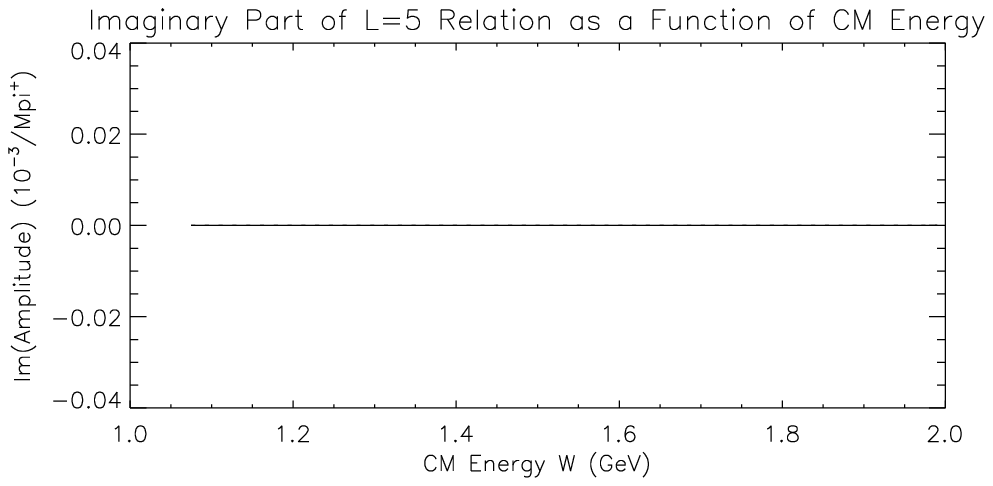}
\caption{
Electric multipole data from the MAID~2003 website~\cite{MAID}.  Solid
lines indicate the l.h.s.\ of relation~(\ref{LOEplus}) for values $L
\! \ge \! 0$, while dotted lines represent the r.h.s.}
\label{LOEpplot}
\end{figure}

We begin with an illustration of the electric multipole results,
Eqs.~(\ref{LOEminus})--(\ref{pi0e}).  In Fig.~\ref{LOEmplot} we plot
the left-hand side (l.h.s.) and right-hand side (r.h.s.) of
Eq.~(\ref{LOEminus}) for $L \! = \! 2$--5, and similarly for
Eq.~(\ref{LOEplus}) with $L \! = \! 0$--5 in Fig.~\ref{LOEpplot}.  It
is immediately clear that the relations are convincing, particularly
in the energy range below resonances---after all, it has been known
for a long time that isoscalar amplitudes are suppressed compared to
isovector amplitudes.  The $1/N_c$ expansion simply provides an
expectation for the relative magnitude of the difference; indeed, the
agreement often seems better than $1/N_c$, or 1 part in 3.

%
%
%

Confronting Eq.~(\ref{pi0e}) with data is more difficult.  One may,
for example, superimpose plots of $M_{L\pm1,L,\pm}^{\textrm{e},\,
p(\pi^0)p}$ with the corresponding charge-exchange amplitudes and ask
whether the former are truly $O(1/N_c)$ smaller than the latter.
Since both of these amplitudes have their own unique structure as
functions of energy, a sort of averaging procedure is necessary, and a
decisive result is not immediately visible.  We therefore do not
include such plots in this work.  Indeed, there are certain energy
regions where the $\pi^0$ electric multipoles are actually larger than
their charged counterparts, particularly for the imaginary parts in
the lower partial waves.  By and large, however, the $\pi^0$ e
amplitudes tend to be smaller at most energies, in general agreement
with Eq.~(\ref{pi0e}).

\begin{figure}[ht]
\epsfxsize 2.5 in \epsfbox{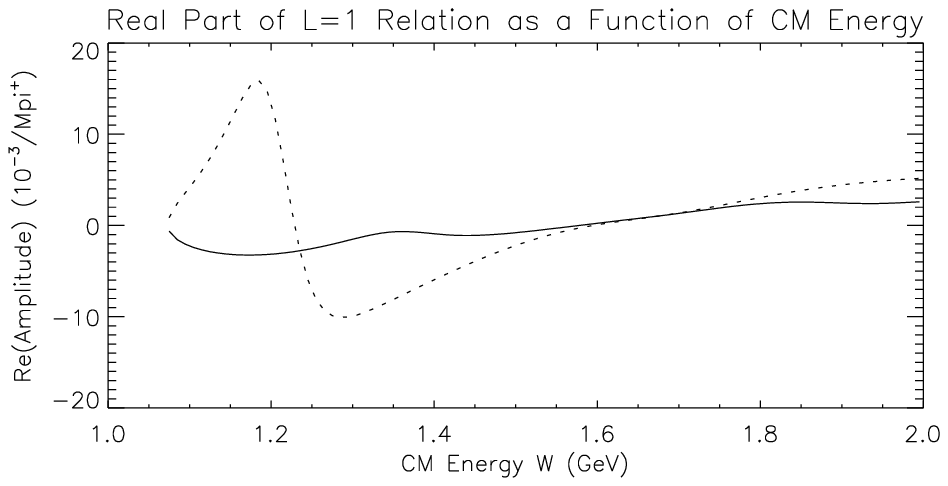} \hspace{5em}
\epsfxsize 2.5 in \epsfbox{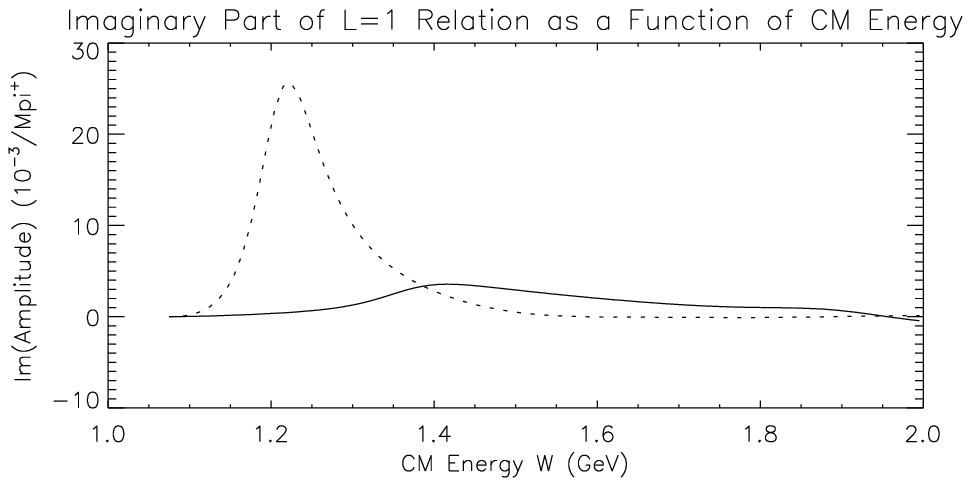} \\
\epsfxsize 2.5 in \epsfbox{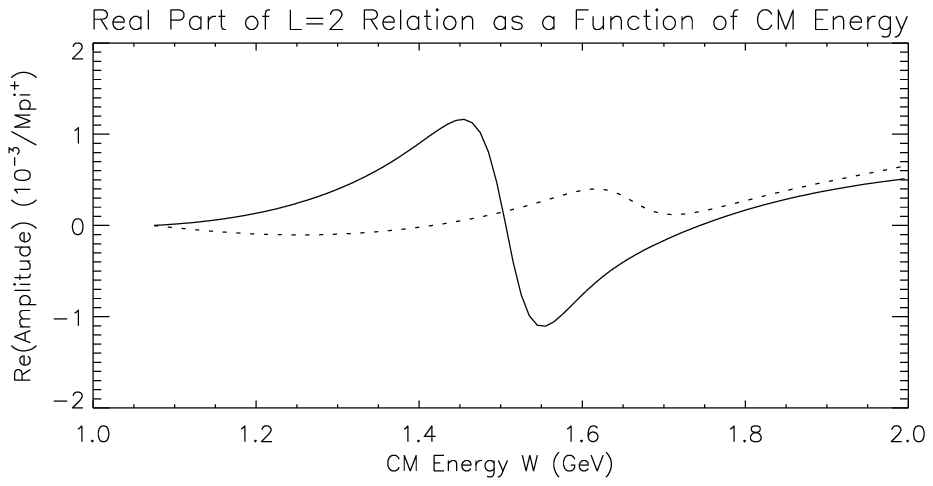} \hspace{5em}
\epsfxsize 2.5 in \epsfbox{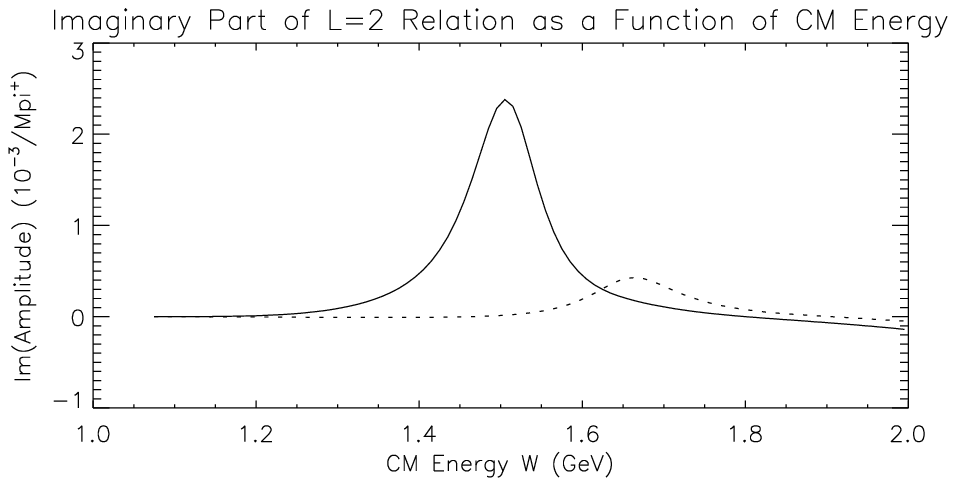} \\
\epsfxsize 2.5 in \epsfbox{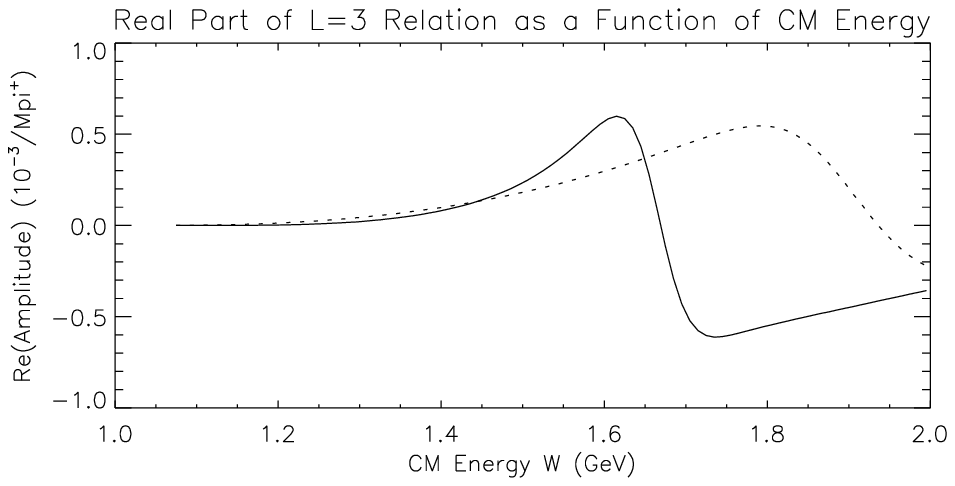} \hspace{5em}
\epsfxsize 2.5 in \epsfbox{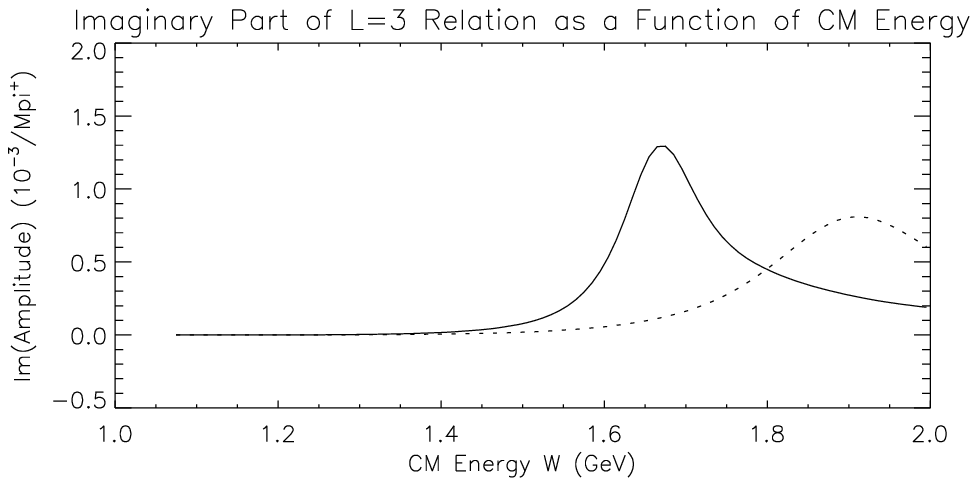} \\
\epsfxsize 2.5 in \epsfbox{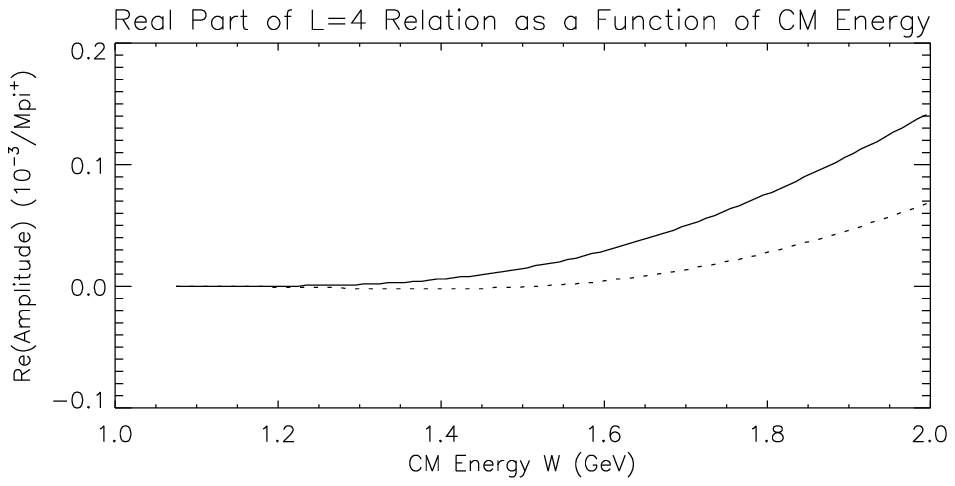} \hspace{5em}
\epsfxsize 2.5 in \epsfbox{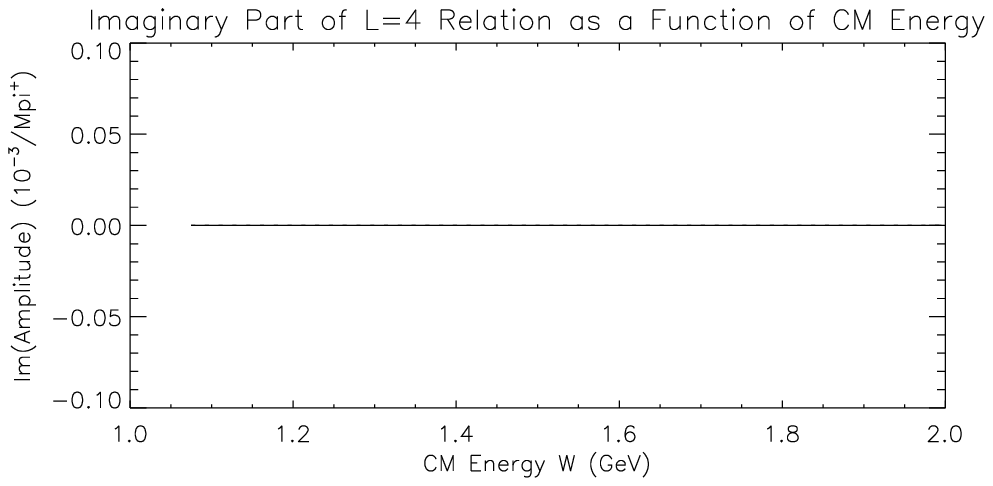} \\
\epsfxsize 2.5 in \epsfbox{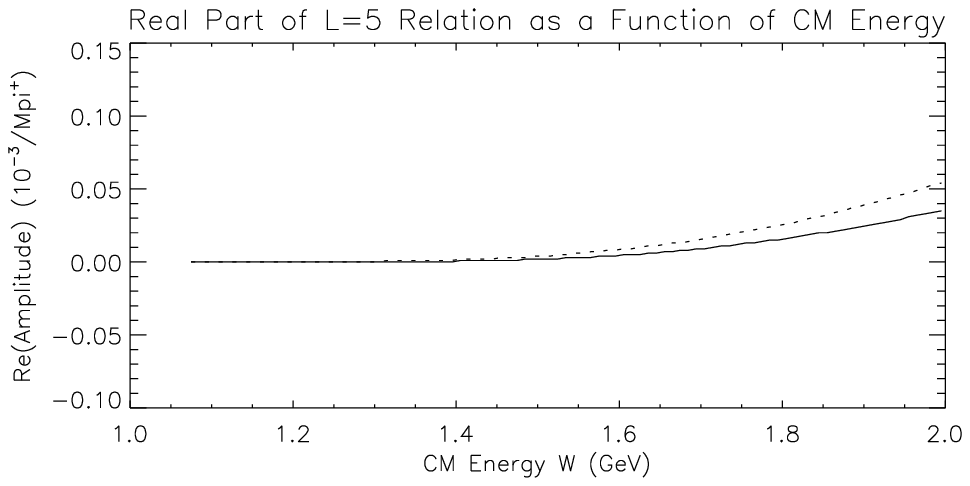} \hspace{5em}
\epsfxsize 2.5 in \epsfbox{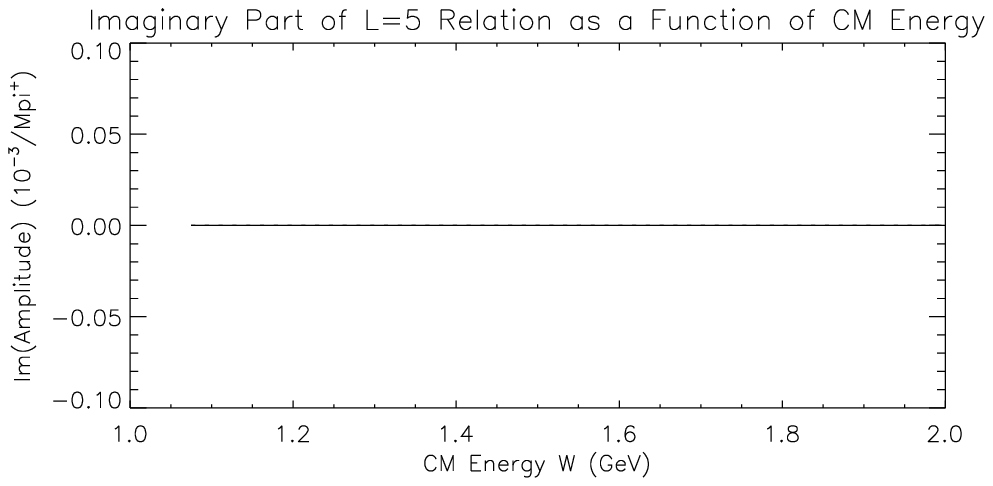}
\caption{
Magnetic multipole data from the MAID~2003 website~\cite{MAID}.  Solid
lines indicate the l.h.s.\ of Eq.~(\ref{LOMnought}) for values $L \!
\ge \! 1$, while dotted lines represent the r.h.s.}
\label{mag0fig}
\end{figure}

We next plot the two sides of the $\pi^0$ magnetic relation
Eq.~(\ref{LOMnought}) in Fig.~\ref{mag0fig} for $L \! = \! 1$--5.
Agreement for the $L \! = \! 1$ partial wave is particularly poor
because of the presence of the $\Delta^+$(P$_{33}$) resonance, which
in the large $N_c$ limit is a stable partner of the nucleons.  As $L$
increases, however, one observes an increasingly satisfactory
comparison.  Even in $L \! = \! 2$, where the resonances D$_{13}
(1520)$ and D$_{15}(1675)$ appear separated and quite different in
amplitude, there is good reason for optimism, as we show below for
on-resonance parameters for the charge-changing amplitudes.

%
%

\begin{figure}[ht]
\epsfxsize 2.5 in \epsfbox{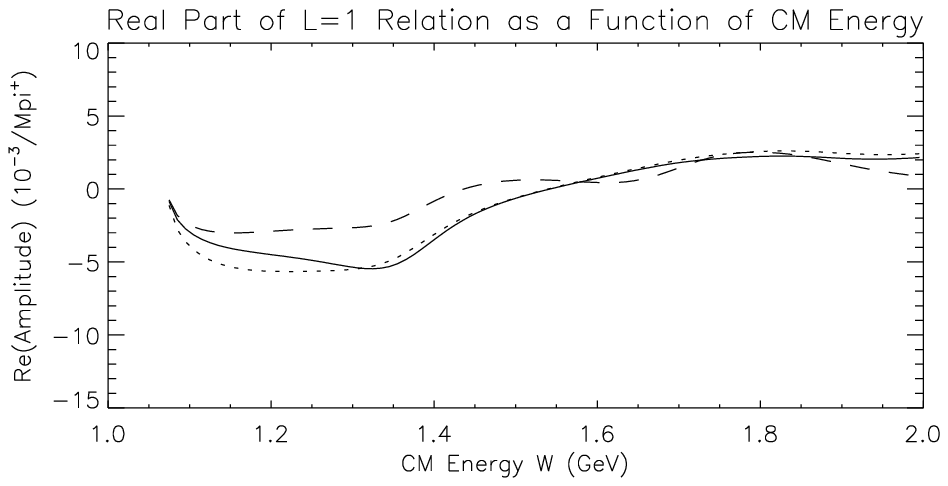} \hspace{5em}
\epsfxsize 2.5 in \epsfbox{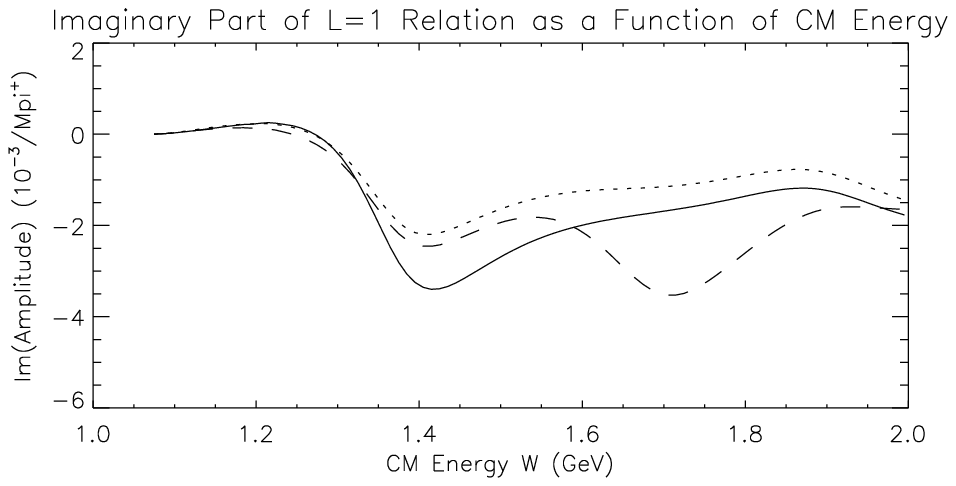} \\
\epsfxsize 2.5 in \epsfbox{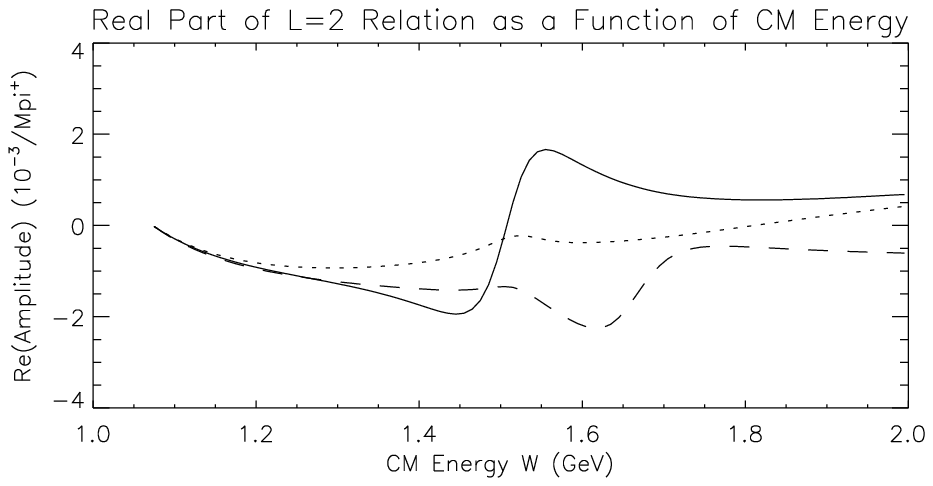} \hspace{5em}
\epsfxsize 2.5 in \epsfbox{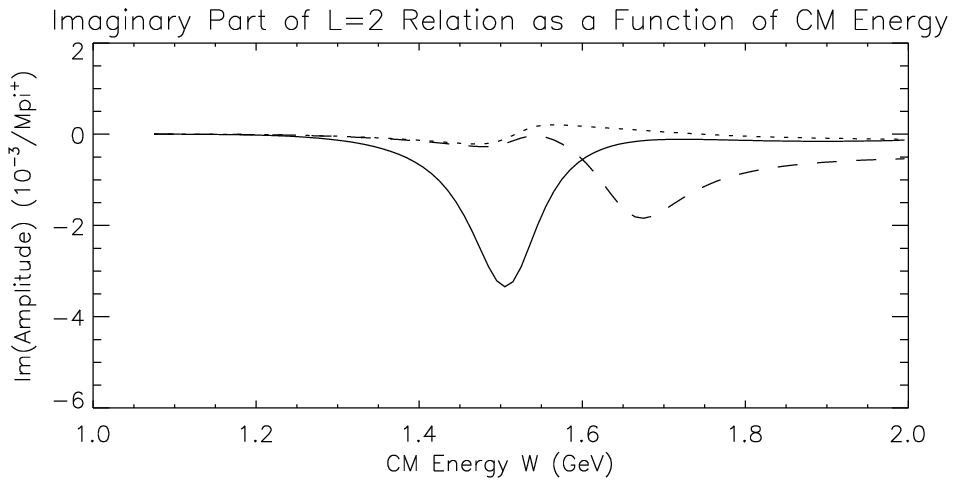} \\
\epsfxsize 2.5 in \epsfbox{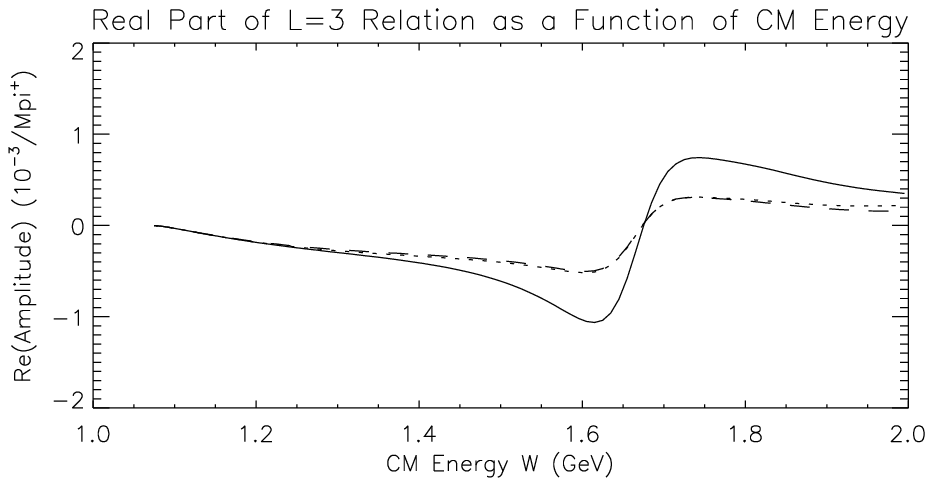} \hspace{5em}
\epsfxsize 2.5 in \epsfbox{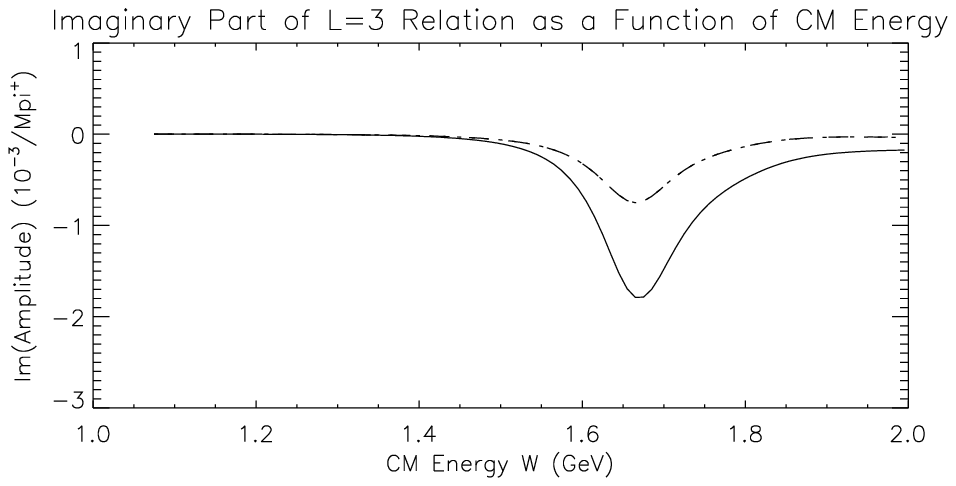} \\
\epsfxsize 2.5 in \epsfbox{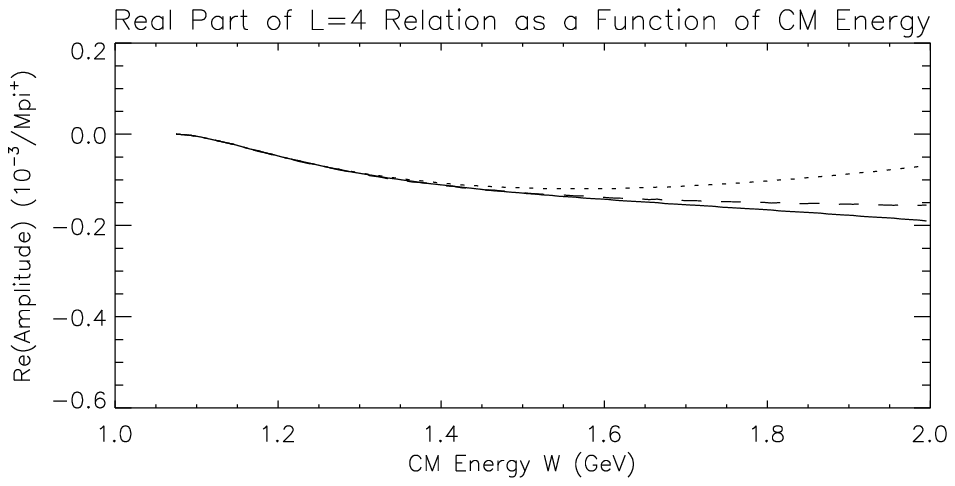} \hspace{5em}
\epsfxsize 2.5 in \epsfbox{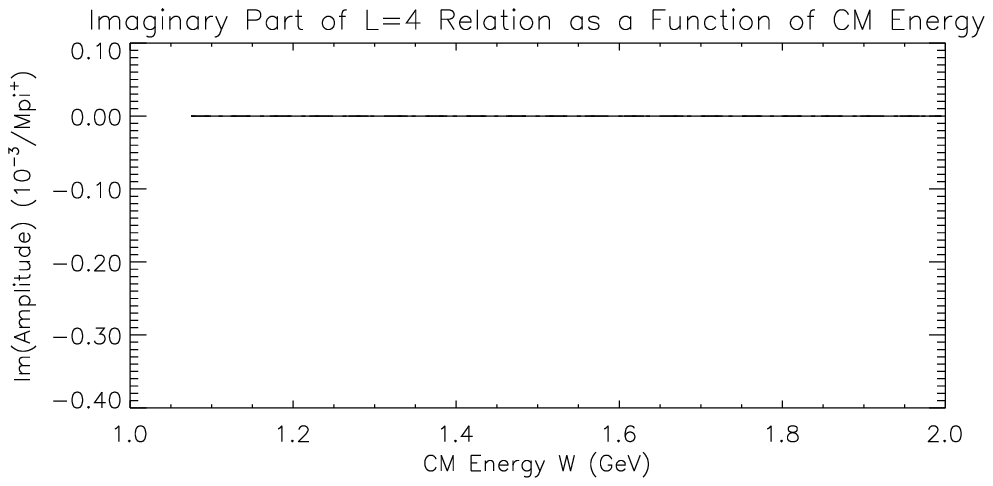} \\
\epsfxsize 2.5 in \epsfbox{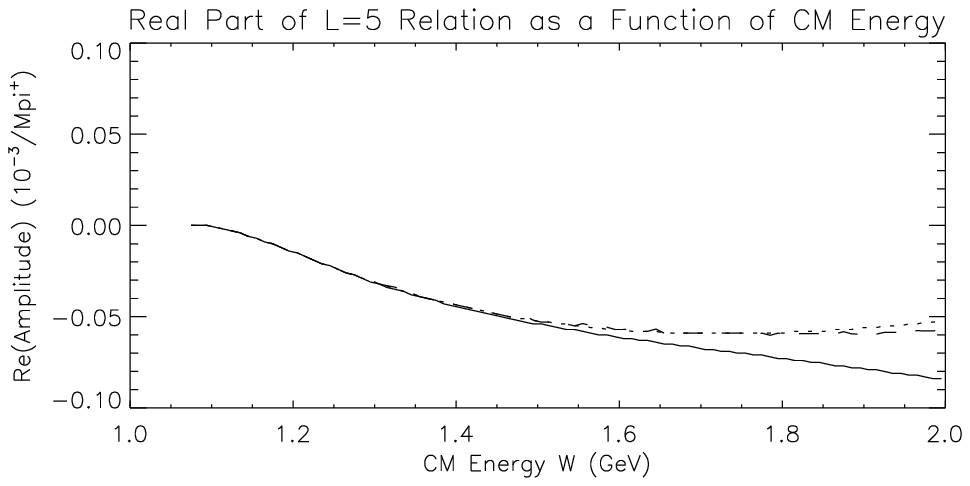} \hspace{5em}
\epsfxsize 2.5 in \epsfbox{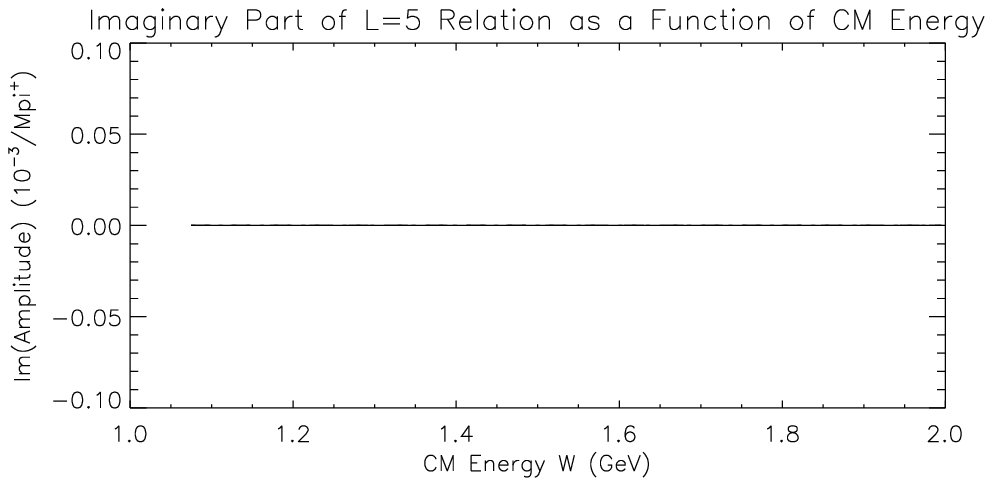}
\caption{
Magnetic multipole data from the MAID~2003 website~\cite{MAID}.  Solid
lines indicate the l.h.s.\ of relations~(\ref{LOMchain}) and
(\ref{NLO}) for values $L \! \ge \! 1$, while dotted lines represent
the LO term [first r.h.s.\ of Eq.~(\ref{LOMchain})], and dashed lines
include the NLO term in Eq.~(\ref{NLO}).}
\label{MAIDmagfig}
\end{figure}

\begin{figure}[ht]
\epsfxsize 2.5 in \epsfbox{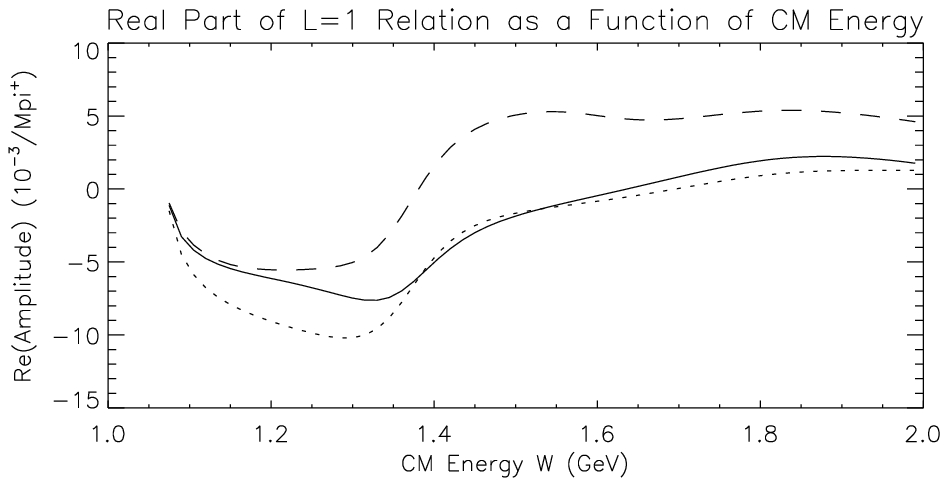} \hspace{5em}
\epsfxsize 2.5 in \epsfbox{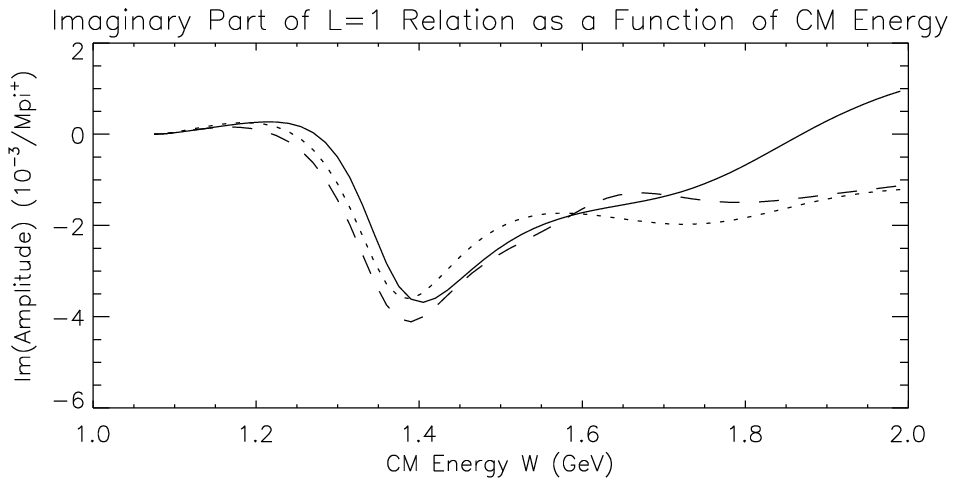} \\
\epsfxsize 2.5 in \epsfbox{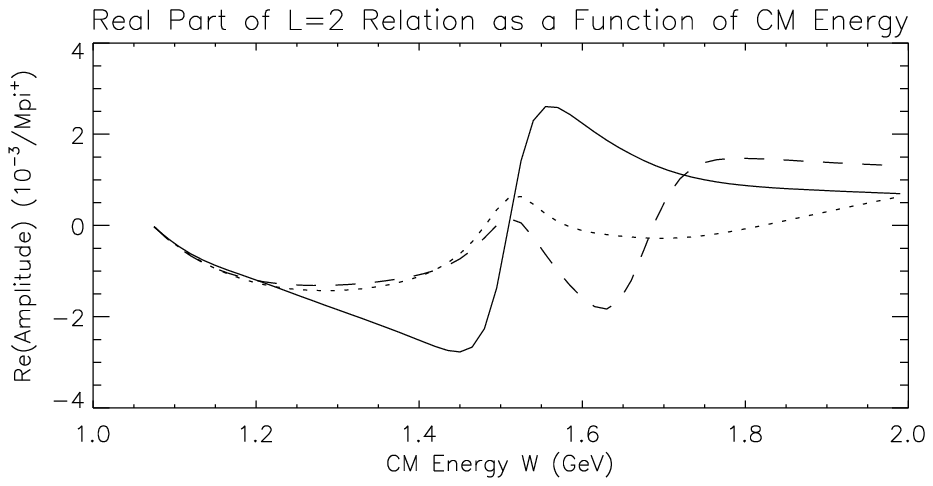} \hspace{5em}
\epsfxsize 2.5 in \epsfbox{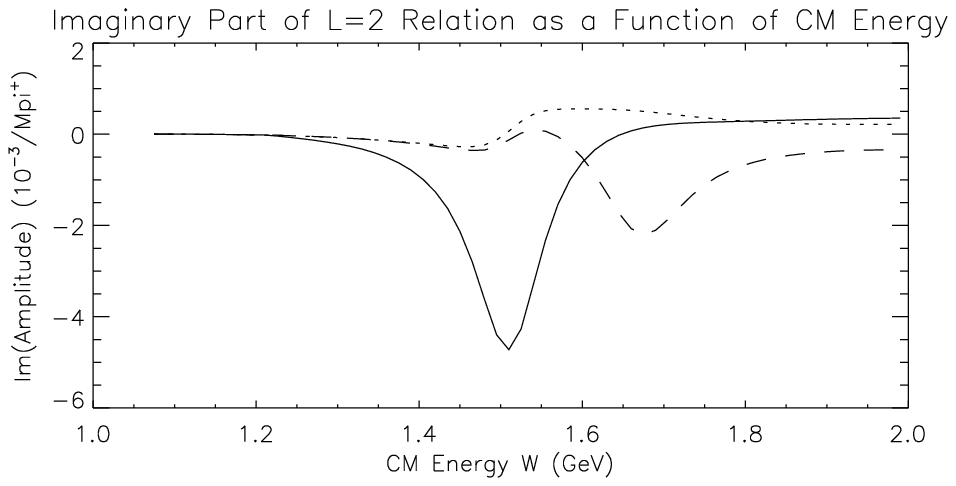} \\
\epsfxsize 2.5 in \epsfbox{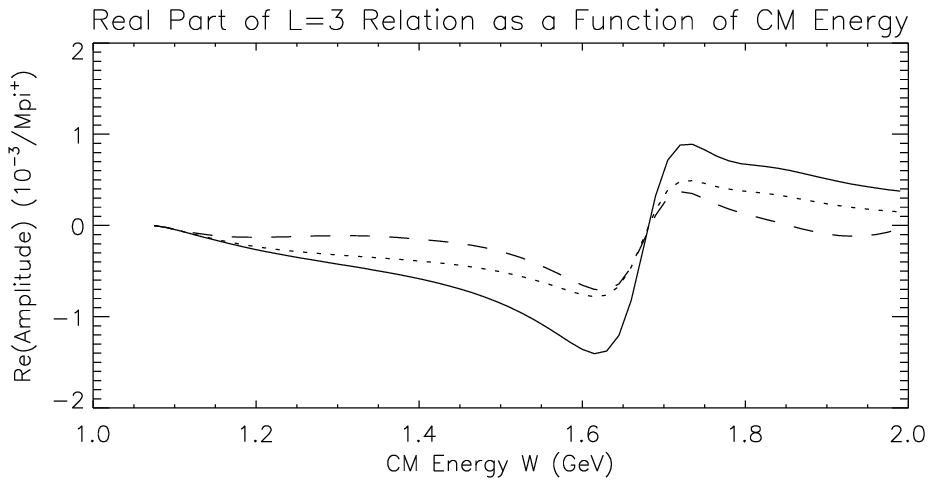} \hspace{5em}
\epsfxsize 2.5 in \epsfbox{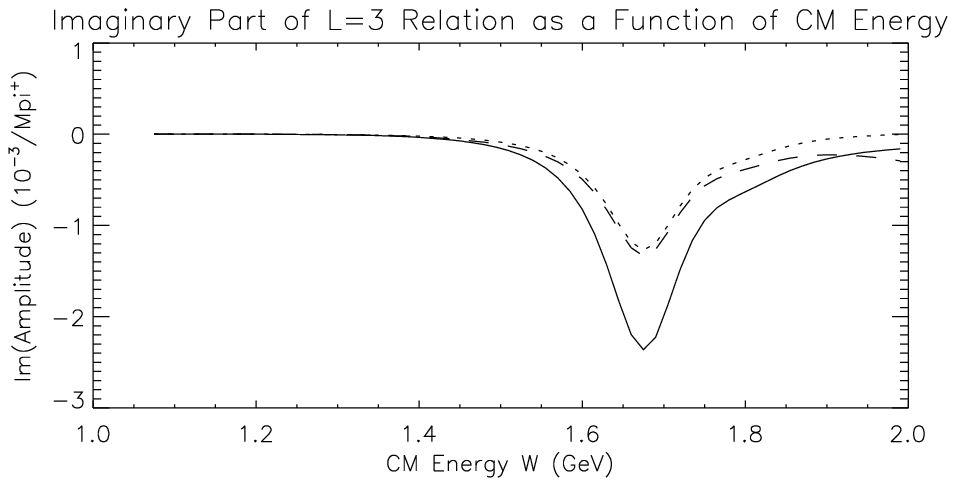} \\
\epsfxsize 2.5 in \epsfbox{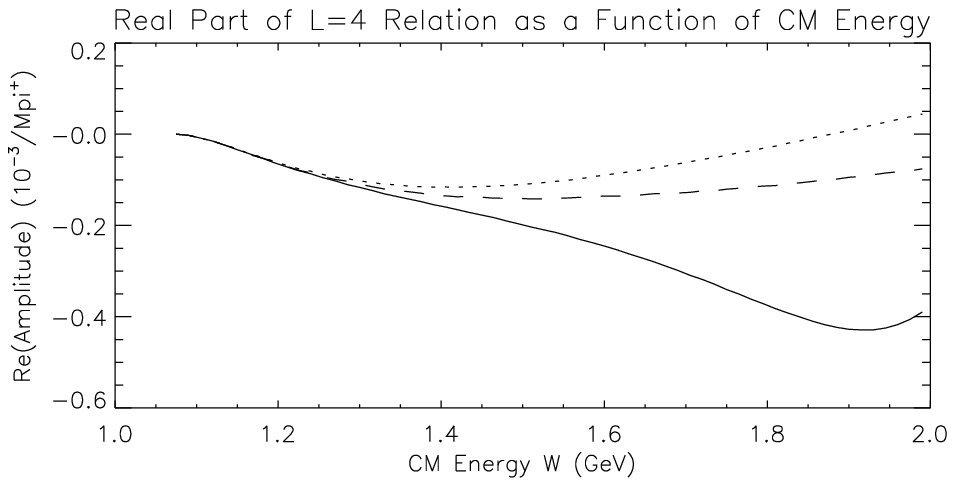} \hspace{5em}
\epsfxsize 2.5 in \epsfbox{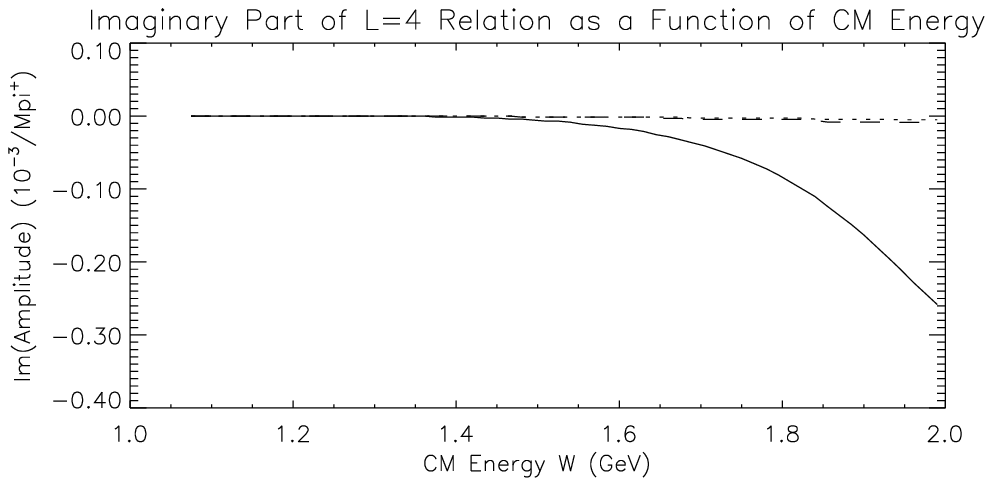} \\
\epsfxsize 2.5 in \epsfbox{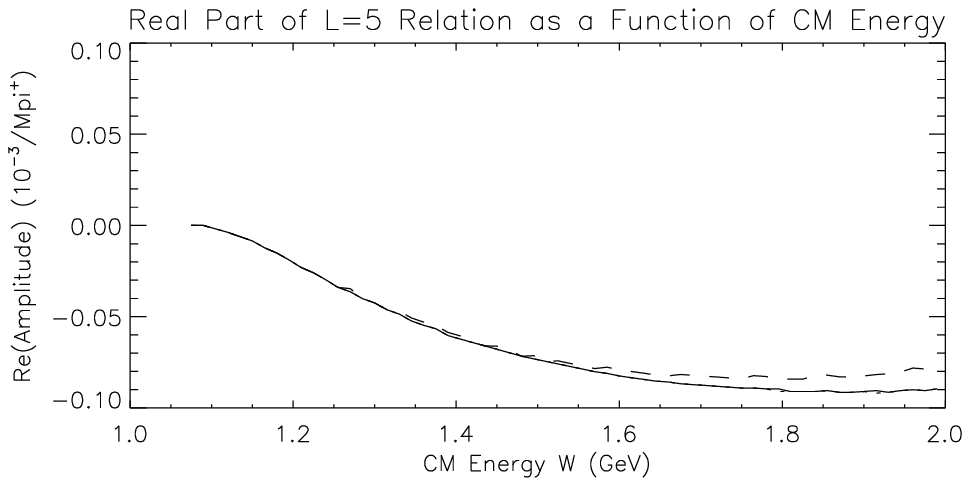} \hspace{5em}
\epsfxsize 2.5 in \epsfbox{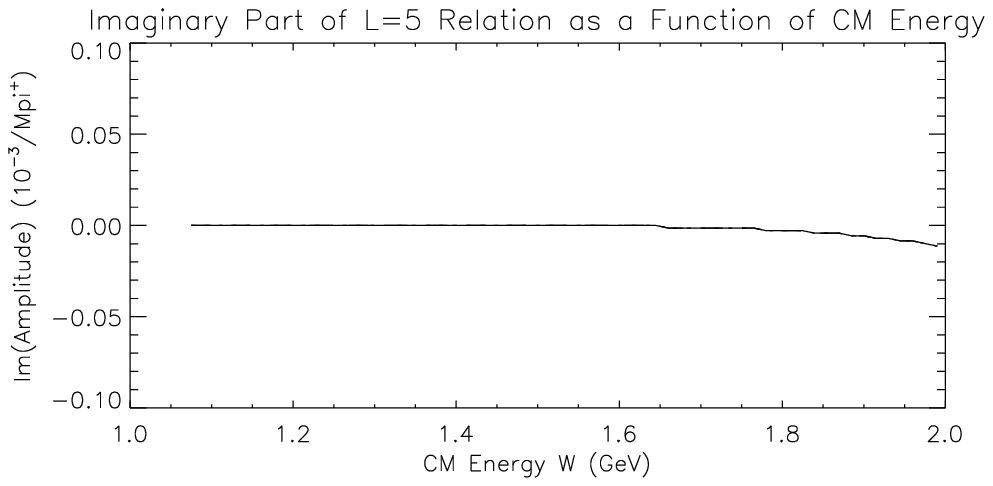}
\caption{
Magnetic multipole data from the SAID website~\cite{SAID}.  Again,
solid lines indicate the l.h.s.\ of the relations~(\ref{LOMchain}) and
(\ref{NLO}) for $L \! \ge \! 1$, while the dotted lines represent the
LO term [first r.h.s.\ of Eq.~(\ref{LOMchain})], and dashed lines
include the NLO term in Eq.~(\ref{NLO}).}
\label{SAIDmagfig}
\end{figure}

Turning now to the charged magnetic multipole relations, we
simultaneously test both LO [leftmost of Eq.~(\ref{LOMchain})] and NLO
[Eq.~(\ref{NLO})] relations in a single set of plots, for $L \! =
\! 1$--5.  In each plot three curves appear, corresponding to
$M^{p(\pi^+)n}_{L-}$ [the l.h.s.\ of Eqs.~(\ref{LOMchain}) and
(\ref{NLO})] and the r.h.s.'s of the LO and NLO relations.  Since NLO
relations are more delicate, we present these combinations using both
MAID (Fig.~\ref{MAIDmagfig}) and SAID (Fig.~\ref{SAIDmagfig}) data.

%
%
%
%
%

One can infer several interesting conclusions from these figures.
First, the LO relations definitely have merit, particularly in the
energy range below the appearance of resonances.  This is true for all
partial waves, real and imaginary parts alike.  However, the addition
of NLO terms does not seem to greatly improve agreement between the
two curves; indeed, in certain low partial waves ({\it e.g.}, $L \! =
\! 1$), the addition of NLO terms seems to make the agreement worse.
However, a clue to what is happening may be gleaned from the fact the
NLO terms in Eq.~(\ref{NLO}) may introduce resonances completely
absent from the LO terms.

These plots reveal the strong effect of resonances in the lower
partial waves on the quality of our predictions.  This was noted
earlier by Schwesinger {\it et al.}~\cite{Report} when they attempted
to compare their Skyrme model relations with experiment.  Rather than
compare the multipole amplitudes along the full energy range, they
proposed an alternate testing method using the resonance couplings
(obtainable through the helicity amplitudes) for the relevant
resonances.

Such an approach is all the more sensible in the $1/N_c$ expansion,
where resonances that would be degenerate in the large $N_c$ limit may
differ in mass by as much as 300~MeV.  For example, the $\Delta$-$N$
mass difference is formally only an $O(1/N_c^1)$ effect.  One should
not be surprised if LO and NLO terms differ by humps that are
shifted with respect to each other.

One can proceed in a similar manner to that of Ref.~\cite{Report},
once Eqs.~(\ref{LOMchain}) and (\ref{NLO}) are written in terms of the
Walker helicity amplitudes~\cite{walker} $A^p$, $A^n$, $B^p$, and
$B^n$, which are, respectively, proportional to the helicity
amplitudes $A^p_{1/2}$, $A^n_{1/2}$, $A^p_{1/2}$, and $A^n_{3/2}$ at
each resonance given in the {\it Review of Particle
Properties}~\cite{PDG}.  In the present case, each of these amplitudes
may have either of $J \!  = \!  L \pm \frac{1}{2}$.  The conversion
between these amplitudes is outlined in the Appendix; the final result
is:
\begin{eqnarray}
\left[A^p - A^n \right]_{L-} +
\frac{1}{2}(L-1)\left[B^p - B^n \right]_{L-}  & = &
O(N_c^{-1}), \label{coupling1} \\
\left[A^p - A^n \right]_{L-} +
\frac{1}{2}(L-1) \left[B^p - B^n \right]_{L-} & & \nonumber \\
+ \left[A^p - A^n \right]_{L+} -
\frac{1}{2}(L+2) \left[B^p - B^n \right]_{L+} & = &
O(N_c^{-2}).\label{coupling2}
\end{eqnarray}

We now consider each partial wave and insert the resonance couplings
for nearby $I \! = \! \frac 1 2$ resonances into the above formulas
(paired $I \! = \! \frac 3 2$ resonances appear to occur too high in
energy to significantly influence these plots).  After consulting
Ref.~\cite{PDG}, one sees that only $L \!  = \! 2$ provides a
meaningful test since D$_{13}$(1520) and D$_{15}$(1675) can be grouped
together as distinct resonances appearing in one of the plotted
partial waves.  This is fortunate, because the $L \! = \! 2$ plot is
the most inconclusive of those discussed above.  Resonances in other
partial waves are either poorly resolved or split too far apart to
make a convincing match.  We evaluate the l.h.s.'s of
Eqs.~(\ref{coupling1}) and (\ref{coupling2}), and also show in curly
braces the sum of the absolute values of each term to demonstrate the
extent of the cancellations.  If the $1/N_c$ expression is working,
the l.h.s.\ in the first should be about 1/3 of the corresponding
factor in braces, and that in the second should be about 1/9.
\begin{eqnarray}
\textrm{l.h.s.} \, (\ref{coupling1}) &=& -38.4 \pm 5.6 \;\; \{ 100.9 \}
\;\times 10^{-3} \,
\textrm{GeV}^{-1}, \label{1st} \\
\textrm{l.h.s.} \, (\ref{coupling2}) &=& -18.2 \pm 8.5 \;\; \{ 140.2 \}
\;\times 10^{-3} \, \textrm{GeV}^{-1} . \label{2nd}
\end{eqnarray}
Expressed as ratios, the results are $-0.38 \pm 0.06$ and $-0.13
\pm 0.06$, respectively.  One sees that the behavior is exactly
what one would expect from the $1/N_c$ expansion.  Indeed, if
anything, the agreement is better than one might expect.  For example,
a central value in Eq.~(\ref{1st}) of 20 or 50 would still be
acceptable.  We conclude that Eq.~(\ref{NLO}) works well, even though
the presence of somewhat separated resonances obscures agreement over
the full energy range.

\section{Conclusions} \label{concl}

We have presented model-independent expressions, derived from the
$1/N_c$ expansion of QCD, for pion photoproduction multipole
amplitudes.  This expansion yields several nontrivial predictions that
can be tested with experimental data.  We find that the relations
holding to leading order in $1/N_c$ match the data quite well in most
cases, particularly in the region between threshold and the onset of
resonances.  The relations holding at next-to-leading order in $1/N_c$
appear to yield a more modest improvement if one insists on
considering the amplitudes at all energy scales, including the
resonant region.  However, when the same relations are employed using
parameters extracted directly from distinct resonances appearing in
partial waves on the two sides of these equations, the agreement with
the expectations of the $1/N_c$ expansion---at both leading and
subleading order---is remarkable.

\section*{Acknowledgments}

D.C.D.\ would like to thank E.~Beise and J.J.~Kelly for explaining
the SAID and MAID data.  The work of T.D.C.\ and D.C.D.\ was
supported in part by the U.S.\ Department of Energy under Grant
No.\ DE-FG02-93ER-40762.  The work of R.F.L.\ and D.R.M.\ was
supported by the National Science Foundation under Grant No.\
PHY-0140362.

\appendix
\section{Helicity Amplitudes}

The magnetic multipoles $M_{L\pm}$ can be rewritten in terms of
helicity amplitudes.  To do this, one introduces the Walker helicity
elements~\cite{walker}, $A_{L\pm}$ and $B_{L\pm}$.  The labels $A$ and
$B$ refer to an initial $\gamma N$ state with total angular momentum
$J \! = \! L \! \pm \! \frac{1}{2}$ that has helicity $\frac{1}{2}$
and $\frac{3}{2}$, respectively.  They are related for $L \! \ge \! 1$
by [Ref.~\cite{walker}, Eq.~(25)]:
\begin{eqnarray}
M_{L+} &=& \frac{1}{L+1}\left[A_{L+} - \frac{1}{2}(L+2)B_{L+}\right],\\
M_{L-} &=& \frac{1}{L}\left[A_{L-} +
\frac{1}{2}(L-1)B_{L-}\right].
\end{eqnarray}
These should be regarded as eight equations, two for each of the four
possible pion photoproduction reactions.  This is sufficient to obtain
Eqs.~(\ref{coupling1}) and (\ref{coupling2}).

The Walker helicity elements can be then be written in terms of
helicity amplitudes $A^p_{1/2}$, $A^p_{3/2}$, $A^n_{1/2}$, and
$A^n_{3/2}$, whose numerical values are tabulated in Ref.~\cite{PDG}.
The subscript indicates the helicity of the state, while the
superscript indicates the initial nucleon.  The relationship between
these two representations is given by Eqs.~(9.8) and (9.9) of
Ref.~\cite{Report}:
\begin{eqnarray}
{\rm Im} A^\beta_{L \pm} & = & \mp f A^\beta_{1/2} \ , \nonumber \\
{\rm Im} B^\beta_{L \pm} & = & \pm f \sqrt{ \frac{16}{(2J-1)(2J+3)} }
A^\beta_{3/2} \ , \nonumber \\
f & = & \sqrt{ \frac{1}{(2J+1) \pi} \frac{k_\gamma}{k_\pi}
\frac{M_N}{M_R} \frac{\Gamma_\pi}{\Gamma^2} } \ ,
\end{eqnarray}
where $\beta$ refers to the initial isospin of the $\gamma N$ system
(and therefore subsumes a Clebsch-Gordan coefficient when one refers
to a specific nucleon charge state). $k_\gamma$ and $k_\pi$ are the
c.m.\ 3-momenta of the photon and pion, respectively. $M_N$ and $M_R$
are the nucleon and resonance masses, and $\Gamma_\pi$ and $\Gamma$
are the pionic and total widths of the resonance, respectively.


\begin{thebibliography}{99}

\bibitem{CDLN}
T.D.~Cohen, D.C.~Dakin, R.F.~Lebed, and A.~Nellore, Phys.\ Rev.\ D
{\bf 70}, 056004 (2004).

\bibitem{GS}
J.-L.~Gervais and B.~Sakita, Phys.\ Rev.\ Lett.\ {\bf 52}, 87
(1984); Phys.\ Rev.\ D {\bf 30}, 1795 (1984).

\bibitem{DM}
R.F.~Dashen and A.V.~Manohar, Phys.\ Lett.\ B {\bf 315}, 425
(1993); {\bf 315}, 438 (1993).

\bibitem{DJM}
R.F.~Dashen, E.~Jenkins, and A.V.~Manohar, Phys.\ Rev.\ D {\bf
49}, 4713 (1994).

\bibitem{CL}
T.D.~Cohen and R.F.~Lebed, Phys.\ Rev.\ Lett.\ {\bf 91}, 012001
(2003); Phys.\ Rev.\ D {\bf 67}, 096008 (2003); Phys.\ Rev.\ D
{\bf 68}, 056003 (2003).

\bibitem{lead}
The leading order in absolute terms is $O(e/\sqrt{N_c})$, but this is
irrelevant for our calculations.

\bibitem{ES}
C.~Eckart and B.~Schwesinger, Nucl.\ Phys.\ {\bf A458}, 620
(1986).

\bibitem{Report}
B.~Schwesinger, H.~Weigel, G.~Holzwarth, and A.~Hayashi, Phys.\
Reports {\bf 173}, 173 (1989).

\bibitem{EW}
E.~Witten, Nucl.\ Phys.\ {\bf B160}, 57 (1979).

\bibitem{ANW}
G.~Adkins, C.~Nappi, and E.~Witten, Nucl.\ Phys.\ {\bf B228}, 552
(1983).

\bibitem{CLpenta}
T.D.~Cohen and R.F.~Lebed, Phys.\ Lett.\ B {\bf 578}, 150 (2004).

\bibitem{HEHWMK}
A.~Hayashi, G.~Eckart, G.~Holzwarth, and H.~Walliser, Phys.\ Lett.\
{\bf B147}, 5 (1984); M.P.~Mattis and M.~Karliner , Phys.\ Rev.\ D
{\bf 31}, 2833 (1985).

\bibitem{Mat}
M.P.~Mattis and M.E.~Peskin, Phys.\ Rev.\ D {\bf 32}, 58 (1985);
M.P.~Mattis, Phys.\ Rev.\ Lett.\ {\bf 56}, 1103 (1986); Phys.\ Rev.\ D
{\bf 39}, 994 (1989); Phys.\ Rev.\ Lett.\ {\bf 63}, 1455 (1989).

\bibitem{MM}
M.P.~Mattis and M.~Mukerjee, Phys.\ Rev.\ Lett.\ {\bf 61}, 1344
(1988).

\bibitem{BLP_RQT}
V.B.~Berestetskii, E.M.~Lifshitz, and L.P.~Pitaevskii, {\it
Relativistic Quantum Theory}, \\ (Addison-Wesley Publishing, Reading,
Massachusetts, 1971).

\bibitem{BL}
A.J.~Buchmann and R.F.~Lebed, Phys.\ Rev.\ D {\bf 62}, 096005 (2000).

\bibitem{BHL}
A.J.~Buchmann, J.A.~Hester, and R.F.~Lebed, Phys.\ Rev.\ D {\bf
66}, 056002 (2002).

\bibitem{JJMLM}
E.~Jenkins, X.~Ji, and A.~Manohar, Phys.\ Rev.\ Lett.\ {\bf 89},
242001 (2002); R.F.~Lebed and D.R.~Martin, Phys.\ Rev.\ D {\bf 70},
016008 (2004).

\bibitem{KapMan}
D.B.~Kaplan and A.V.~Manohar, Phys.\ Rev.\ C {\bf 56}, 76 (1997).

\bibitem{edmonds}
A.R.~Edmonds, {\it Angular Momentum in Quantum Mechanics}
(Princeton Univ.\ Press, Princeton, NJ, 1996).

\bibitem{neut}
Of course, this is an empirically testable hypothesis on such data as
exists.

\bibitem{SAID}
The SAID data is available at George Washington University's
Center for Nuclear Studies website:  http://gwdac.phys.gwu.edu.

\bibitem{MAID}
The MAID data is available at the Universit\"{a}t Mainz Institute for
Nuclear Physics website: http://www.kph.uni-mainz.de/maid/.

\bibitem{walker}
R.L.~Walker, Phys.\ Rev.\ {\bf 182}, 1729 (1969).

\bibitem{PDG}
Particle Data Group, Phys.\ Lett.\ B {\bf 592}, 1 (2004).

\end{thebibliography}
\end{document}